\documentclass[twoside,twocolumn,9pt]{article}
\usepackage{extsizes}
\usepackage[super,sort&compress,comma]{natbib} 
\usepackage[version=3]{mhchem}
\usepackage[left=1.5cm, right=1.5cm, top=1.785cm, bottom=2.0cm]{geometry}
\usepackage{balance}
\usepackage{times,mathptmx}
\usepackage{sectsty}
\usepackage{graphicx} 
\usepackage{lastpage}
\usepackage[format=plain,justification=justified,singlelinecheck=false,font={stretch=1.125,small,sf},labelfont=bf,labelsep=space]{caption}
\usepackage{float}
\usepackage{fancyhdr}
\usepackage{fnpos}
\usepackage[english]{babel}
\usepackage{array}
\usepackage{droidsans}
\usepackage{charter}
\usepackage[T1]{fontenc}
\usepackage[usenames,dvipsnames]{xcolor}
\usepackage{setspace}
\usepackage[compact]{titlesec}
\usepackage{color}
\usepackage{amsmath,amssymb,stmaryrd,bm}
\usepackage[utf8]{inputenc}

\definecolor{cream}{RGB}{222,217,201}

\newcommand{\dd}{\ensuremath{\, \mathrm{d}}}
\newcommand{\vv}[1]{\ensuremath{\mathbf{#1}}}

\begin{document}

\pagestyle{fancy}
\thispagestyle{plain}
\fancypagestyle{plain}{

\fancyhead[C]{\includegraphics[width=18.5cm]{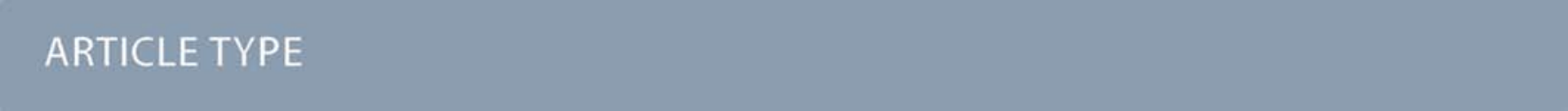}}
\fancyhead[L]{\hspace{0cm}\vspace{1.5cm}\includegraphics[height=30pt]{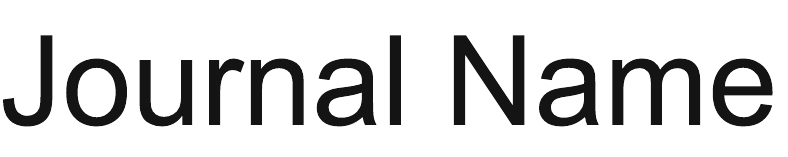}}
\fancyhead[R]{\hspace{0cm}\vspace{1.7cm}\includegraphics[height=55pt]{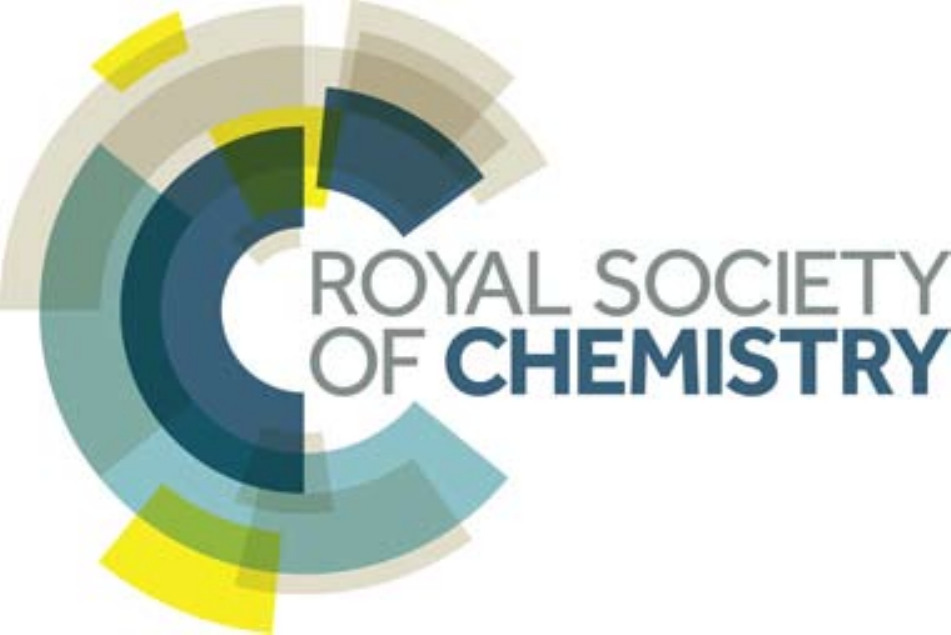}}
\renewcommand{\headrulewidth}{0pt}
}

\makeFNbottom
\makeatletter
\renewcommand\LARGE{\@setfontsize\LARGE{15pt}{17}}
\renewcommand\Large{\@setfontsize\Large{12pt}{14}}
\renewcommand\large{\@setfontsize\large{10pt}{12}}
\renewcommand\footnotesize{\@setfontsize\footnotesize{7pt}{10}}
\makeatother

\renewcommand{\thefootnote}{\fnsymbol{footnote}}
\renewcommand\footnoterule{\vspace*{1pt}%
\color{cream}\hrule width 3.5in height 0.4pt \color{black}\vspace*{5pt}} 
\setcounter{secnumdepth}{5}

\makeatletter 
\renewcommand\@biblabel[1]{#1}            
\renewcommand\@makefntext[1]%
{\noindent\makebox[0pt][r]{\@thefnmark\,}#1}
\makeatother 
\renewcommand{\figurename}{\small{Fig.}~}
\sectionfont{\sffamily\Large}
\subsectionfont{\normalsize}
\subsubsectionfont{\bf}
\setstretch{1.125} 
\setlength{\skip\footins}{0.8cm}
\setlength{\footnotesep}{0.25cm}
\setlength{\jot}{10pt}
\titlespacing*{\section}{0pt}{4pt}{4pt}
\titlespacing*{\subsection}{0pt}{15pt}{1pt}

\fancyfoot{}
\fancyfoot[LO,RE]{\vspace{-7.1pt}\includegraphics[height=9pt]{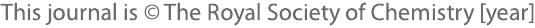}}
\fancyfoot[CO]{\vspace{-7.1pt}\hspace{13.2cm}\includegraphics{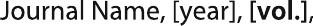}}
\fancyfoot[CE]{\vspace{-7.2pt}\hspace{-14.2cm}\includegraphics{RF}}
\fancyfoot[RO]{\footnotesize{\sffamily{1--\pageref{LastPage} ~\textbar  \hspace{2pt}\thepage}}}
\fancyfoot[LE]{\footnotesize{\sffamily{\thepage~\textbar\hspace{3.45cm} 1--\pageref{LastPage}}}}
\fancyhead{}
\renewcommand{\headrulewidth}{0pt} 
\renewcommand{\footrulewidth}{0pt}
\setlength{\arrayrulewidth}{1pt}
\setlength{\columnsep}{6.5mm}
\setlength\bibsep{1pt}

\makeatletter 
\newlength{\figrulesep} 
\setlength{\figrulesep}{0.5\textfloatsep} 

\newcommand{\topfigrule}{\vspace*{-1pt}%
\noindent{\color{cream}\rule[-\figrulesep]{\columnwidth}{1.5pt}} }

\newcommand{\botfigrule}{\vspace*{-2pt}%
\noindent{\color{cream}\rule[\figrulesep]{\columnwidth}{1.5pt}} }

\newcommand{\dblfigrule}{\vspace*{-1pt}%
\noindent{\color{cream}\rule[-\figrulesep]{\textwidth}{1.5pt}} }

\makeatother

\twocolumn[
  \begin{@twocolumnfalse}
\vspace{3cm}
\sffamily
\begin{tabular}{m{4.5cm} p{13.5cm} }

\includegraphics{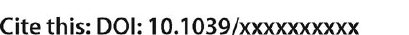} & \noindent\LARGE{\textbf{Phase Separation and Coexistence of Hydrodynamically Interacting Microswimmers$^\dag$}} \\
\vspace{0.3cm} & \vspace{0.3cm} \\

 & \noindent\large{Johannes Blaschke,\textit{$^{ad}$} Maurice Maurer,\textit{$^{b}$} Karthik Menon,\textit{$^{a}$} Andreas Z\"ottl,\textit{$^{c}$} and \mbox{Holger} Stark\textit{$^{ae}$}} \\

\includegraphics{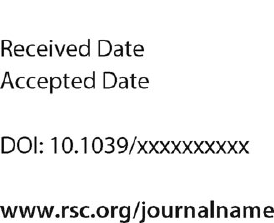} & \noindent\normalsize{A striking feature of the collective behavior of spherical microswimmers is that for sufficiently strong self-propulsion they phase-separate into a dense cluster coexisting with a low-density disordered surrounding.
Extending our previous work, we use the squirmer as a model swimmer and the particle-based simulation method of multi-particle collision dynamics to explore the influence of hydrodynamics on their phase behavior in a quasi-two-dimensional geometry.
The coarsening dynamics towards the phase-separated state is diffusive in an intermediate time regime followed by a final ballistic compactification of the dense cluster.
We determine the binodal lines in a phase diagram of P\'eclet number versus density.
Interestingly, the gas binodals are shifted to smaller densities for increasing mean density or dense-cluster size, which we explain using a recently introduced pressure balance [S. C. Takatori et al., \emph{Phys. Rev. Lett.} \textbf{113}, 028103 (2014)] extended by a hydrodynamic contribution. 
Furthermore, we find that for pushers and pullers the binodal line is shifted to larger P\'eclet numbers compared to neutral squirmers. 
Finally, when lowering the P\'eclet number, the dense phase transforms from a hexagonal ``solid'' to a disordered ``fluid'' state.} \\

\end{tabular}

 \end{@twocolumnfalse} \vspace{0.6cm}

  ]

\renewcommand*\rmdefault{bch}\normalfont\upshape
\rmfamily
\section*{}
\vspace{-1cm}


\footnotetext{\textit{$^{a}$~Institute of Theoretical Physics, Technische Universit\"at Berlin,
Hardenbergstr. 36, D-10623 Berlin, Germany}}
\footnotetext{\textit{$^{b}$~Department of Physics and Astronomy, University of California, Los Angeles, California 90095, USA}}
\footnotetext{\textit{$^{c}$~The Rudolf Peierls Centre for Theoretical Physics, University of Oxford, 
1 Keble Road, Oxford, OX1 3NP, UK}}
\footnotetext{\textit{$^{d}$~E-mail: johannes.blaschke@tu-berlin.de}}
\footnotetext{\textit{$^{e}$~E-mail: holger.stark@tu-berlin.de}}


\footnotetext{\dag~Electronic Supplementary Information (ESI) available: 6 video files. See DOI: 10.1039/xxxxxxxxxx/}





\section{Introduction}
Collective dynamics due to the active motion of microorganisms and artificial microswimmers 
has received a lot of attention among physicists \cite{Zottl:2016kr,Herminghaus:2014cm,Marchetti:2013bp,Saintillan:2013ct,Aranson:2013jt,Kapral:2013fg} as it is relevant both to real world applications \cite{Herminghaus:2014cm,Kapral:2013fg,Palacci:2013eu,Theurkauff:2012jo} as well as for posing fundamental questions in non-equilibrium statistical physics \cite{Takatori:2014do,Bialke:2015cq,Speck:2015gc,Schaar:2015kp,Wittkowski:2014dt,Saintillan:2013ct}. 
From the perspective of physics, a unifying feature of microorganisms and artificial microswimmers is that they propel themselves autonomously through a surrounding fluid. As they are constantly driven out of equilibrium, understanding the collective properties of systems comprised of many swimmers is an important field of ongoing research in statistical physics.

One of the most striking features of swimmer systems is motility-induced phase separation \cite{Cates:2015ft}, where stable clusters of swimmers form due to their active motion alone and without any attractive forces. 
The question of motility-induced phase separation has been treated theoretically using so-called active Brownian particles, which perform a persistent random walk and interact sterically only \cite{Stenhammar:2015ex,Solon:2015hz,Speck:2015gc,Speck:2014iy,Bialke:2013gw,Tailleur:2008kd}.
However, real microswimmers such as ciliated microorganisms \cite{Drescher:2009cy}, 
catalytic Janus particles \cite{Simmchen:1iy,Buttinoni:2013de,Palacci:2013eu,Theurkauff:2012jo,Palacci:2010hk}, or active emulsion droplets 
\cite{Thutupalli:2011bv,Schmitt:2013jo,Herminghaus:2014cm,Maass:2016hz,Schmitt:2016kv} employ propulsion mechanisms reliant on hydrodynamics.
Thus nearby microswimmers can affect each other's velocity and orientation, which can potentially alter their collective behaviour depending on the details of their hydrodynamic interactions \cite{Zottl:2014fn}.

Biological microswimmers employ non-reciprocal cell-body deformations or the collective dynamics of flagella or cilia in order to propel themselves \cite{Purcell:1977tka}, whereas active colloids and droplets move forward by creating a slip velocity close to their surface \cite{Zottl:2016kr}.
The essential features of both of these self-propulsion mechanisms are captured by the so-called ``squirmer'' model \cite{Lighthill:1952hx,Blake:1971ig,Stone:1996ix,ISHIKAWA:2006hf,Downton:2009ig}, which we shall use to generate our hydrodynamic propulsion.

For phase separating systems, it is common to construct the binodal line in order to determine the phase coexistence region.
While there have been many studies quantifying the phase-coexistence regime for active Brownian particles, \cite{Redner:2013jo,Stenhammar:2014fc,Wittkowski:2014dt,Wysocki:2014ky,Yang:2014bl,Speck:2015gc,Stenhammar:2015ex} to list just a few of them, none of these have considered hydrodynamic interactions between the swimmers.
Several recent works studied the collective dynamics of hydrodynamic swimmers \cite{Alarcon:2013dg,Brotto:2013fj,Hennes:2014cw,Li:2014bu,Oyama:2016eh}, but did not look at phase separation.

In our previous study, we examined the collective dynamics and clustering of hydrodynamic squirmers in a quasi-two-dimensional geometry using the method of multi-particle collision dynamics (MPCD) to explicitly simulate the fluid environment \cite{Zottl:2014fn}.
Due to computational limitations, we were restricted to 208 swimmers, which was insufficient to quantify the first-order phase-separation transition.
Hence, we built on this study and extended the simulations to thousands of squirmers by implementing a parallelized version on 1008 processor cores. 
This allows us to investigate the dynamics of a sufficiently large number of squirmers in order to quantitatively resolve phase separation into a dilute gas and a dense phase.

In the following we show phase-separated squirmer systems and explain how we analyze them to obtain the area fractions of the dilute and dense phases. 
After illustrating the coarsening dynamics towards phase separation, we quantify the coexistence region bounded by the binodal line in a diagram P\'eclet number versus density.
Interestingly, the binodal lines differ for varying mean density. We explain this phenomenon as a true hydrodynamic signature using a pressure balance.
In particular, we use the swim pressure introduced in Ref.\ \cite{Takatori:2014do} and extend the approach of Refs.\ \cite{Takatori:2014do,Bialke:2015cq} by including a hydrodynamic pressure generated by squirmers 
at the rim of the dense cluster. 
The binodal line is shifted towards larger P\'eclet numbers, when the force-dipole contribution of the squirmer, characterizing pushers and pullers, is increased. 
Finally, moving the dense phase from large P\'eclet numbers to the critical value, one observes a melting transition between a hexagonal and a fluid phase. 
We start with summarizing computational details.

\section{System and Methods}

Our system consists for $N$ spherical squirmers of radius $R$, where $N$ varies from 3328 to 4624 depending on mean area fraction or density $\bar{\phi}$.
Squirmers propel themselves by applying an axisymmetric surface-velocity field \cite{ISHIKAWA:2006hf,Downton:2009ig},
\begin{equation}
	\vv{v}_s = B_1(1+\beta\hat{\vv{e}}\cdot\hat{\vv{r}}_s)\left[(\hat{\vv{e}}\cdot\hat{\vv{r}}_s)
	\hat{\vv{r}}_s
	-\hat{\vv{e}}\right] \,
	\label{eq:vs-squirmer}
\end{equation}
where $\vv{v}_s$ is the slip velocity, $\vv{r}_s$ points from the squirmer centre to its surface, $\hat{\vv{r}}_s$ is the unit vector along $\vv{r}_s$, and $\hat{\vv{e}}$ is the squirmer's orientation. 
The constant $B_1$ determines the amplitude of the slip velocity and thus its swimming speed $v_0$. 
For a single squirmer in a bulk fluid, $v_0=2B_1/3$. 
We use $B_1$ to control the P\'eclet number, $\mathrm{Pe} = R v_0/D_\mathrm{T}$, as the appropriate thermal diffusion coefficient $D_\mathrm{T}$ is kept constant for all simulations. 
For the MPCD fluid and squirmer parameters presented below, we obtain $\mathrm{Pe} \approx 3546.4 B_1$. 
The additional parameter $\beta$ breaks the symmetry about the squirmer equator. 
For $\beta<0$, the surface field is more concentrated on the southern hemisphere and the squirmer's far field in bulk is that of a pusher.
For $\beta > 0$ a puller is realized and a swimmer with $\beta=0$ is called a  ``neutral squirmer''.

Motivated by experiments \cite{Kruger:2016ju,Herminghaus:2014cm,Thutupalli:2011bv} and following our previous study \cite{Zottl:2014fn}, the swimmers are bounded by two parallel flat plates. 
As in Ref. \cite{Zottl:2014fn} we choose strong confinement with the distance between the plates being $H = 8R/3$.
We implement the full hydrodynamic interactions between squirmers and confining walls using the MPCD simulation technique \cite{Noguchi:2007fia,Padding:2006fo}.
The fluid is represented by of the order of $10^7$ point-like effective fluid particles with mass $m_0$, velocity $\vv{v}_i(t)$, and position $\vv{r}_i(t)$. 
The velocities $\vv{v}_i(t)$ are thermostatted to have a temperature $T_0$.
The fluid particles perform alternating streaming and collision steps. 
During the streaming step, fluid particles move ballistically over a time $\Delta t$ without interacting with each other and reach the position
\begin{equation}
	\vv{r}_i(t+\Delta t) = \vv{r}_i + \vv{v}_i(t)\Delta t \, .
\end{equation}
They do interact with confining boundaries and squirmers, transferring linear and angular momentum.
Interactions with the confining boundaries implement no-slip boundary conditions via the so-called bounce-back rule\cite{Zoettl:2014ev}, while interactions with squirmers generate the surface field of Eq.~\eqref{eq:vs-squirmer}.
For each collision step, particles are sorted into a cubic grid of cell length $a_0$ and with a random offset for each new step.
The average velocity $\bar{\vv{v}} = \langle \vv{v}_i \rangle_\mathrm{cell}$ and centre of mass $\bar{\vv{r}} = \langle \vv{r}_i \rangle_\mathrm{cell}$ of the particles in each cell is determined.
Within each cell they ``collide'' according to the MPCD-AT+a rule\cite{Noguchi:2007fia}, which is reminiscent of an Andersen thermostat and also conserves linear and angular momentum.
Random velocities increments $\delta \vv{v}_i$ for each particle are sampled from a Gaussian distribution with variance $k_\mathrm{B}T_0/m_0$.
The velocity increments would give an overall change in momentum $m_0 \Delta \bar{\vv{v}} = m_0 \langle \delta\vv{v}_i \rangle_\mathrm{cell}$ and angular momentum 
$\Delta\vv{L} = m_0 \langle (\vv{r}_i-\bar{\vv{r}}) \times \delta\vv{v}_i \rangle_\mathrm{cell}$.
Hence, we subtract these changes in the velocity increment for the collision step,
\begin{equation}
	\vv{v}_i(t+\Delta t) = \bar{\vv{v}}(t) + \delta\vv{v}_i - \Delta\bar{\vv{v}} -(\vv{r}_i(t)-\bar{\vv{r}}(t)) \times \vv{I}^{-1} \Delta\vv{L} \, ,
\end{equation}
where $\vv{I}^{-1}$ is the inverse of the moment of inertia tensor for the point-mass distribution of the particles in the cell's centre-of-mass frame.
Thus, the motion of the fluid particles on the length scale of the grid corresponds to solutions of the Navier-Stokes equations \cite{Noguchi:2008bm,Noguchi:2007fia,Gotze:2007ec} including thermal noise.
Furthermore, this scheme accurately reproduces the hydrodynamic flow field of a squirmer \cite{Downton:2009ig,Gotze:2010jl} and their interactions including near-field lubrication effects
\cite{Gotze:2010jl,Zottl:2014fn}.

\subsection{System Parameters}
\label{sec:fluid_properties}

For all simulations discussed here, the fluid and squirmer mass densities are the same, $\rho = 10 m_0/a_0^3$, and the squirmer radius is $R=3a_0$.
To\-ge\-ther with the streaming-time step $\Delta t = 0.02a_0\sqrt{m_0/k_\mathrm{B}T_0}$, the fluid viscosity becomes $\eta = 16.05\sqrt{m_0k_\mathrm{B}T_0}/a_0^2$ taken from\cite{Noguchi:2008bm,Zottl:2014fn}.
For the bulk fluid and the parameters stated above, the self-diffusivity of a passive colloid is
$D_{\mathrm{T}, B} = k_\mathrm{B} T / 6\pi\eta R = 1.1\times 10^{-3} a_0\sqrt{k_\mathrm{B}T_0/m_0}$.
In the presence of walls, the self-diffusivity becomes a position-dependent tensor \cite{Imperio:2011cj}.
We measure the self-diffusivity at the centre between the walls using the Green-Kubo formula. 
We find that $D_{\mathrm{T}, 2D} \approx 0.5D_{\mathrm{T}, B}$, comparable to the result of Ref.\cite{Imperio:2011cj}.
For a colloid close to the walls this value can decrease to approximately $0.25D_{\mathrm{T}, B}$ due to the enhanced friction.
In the following, we use $D_{\mathrm{T}, 2D}$ for a single particle (particle density $\phi=0$) to define the P\'eclet number,  
\begin{equation}
	\mathrm{Pe} = \frac{R v_0}{D_{\mathrm{T},2D}(\phi=0)} \, .
	\label{eq.Pe}
\end{equation}

The orientation of a squirmer also undergoes rotational diffusion, for which MPCD gives the rotational diffusion coefficient $D_{\mathrm{R}} =k_\mathrm{B} T/8\pi\eta R^3 \approx 10^{-4} \sqrt{k_\mathrm{B}T_0/m_0}/a_0^2$ 
in a bulk fluid, which is also approximately valid in our quasi-two-dimensional geometry.
There, rotational diffusion strongly depends on the density of squirmers due to the flow field they create and increases by as much as a factor of 15 in very dense systems\cite{Zottl:2014fn}.
The rotational diffusion coefficient is used to define the persistence number $\mathrm{Pe}_\mathrm{r} = v_0/R D_\mathrm{R}$, which compares persistence or run length $l=v_0 / D_\mathrm{R}$ to $R$.

To summarise, in the following we work with P\'eclet numbers in the range $[190,355]$, which corresponds to persistence numbers $ \mathrm{Pe}_\mathrm{r}$ in the range $[120,220]$ or typical run lengths of single squirmers, $l = v_0/D_\mathrm{R}$, ranging from $120R$ to $220R$.
For high-density simulations without phase separation, where rotational diffusion is considerably enhanced, we have $\mathrm{Pe}_\mathrm{r} \gtrsim 8$ or $l \gtrsim 8R$.

\section{Results}

\subsection{Phase Separation and Coarsening Dynamics}

\begin{figure} 
	\includegraphics[width=0.5\textwidth]{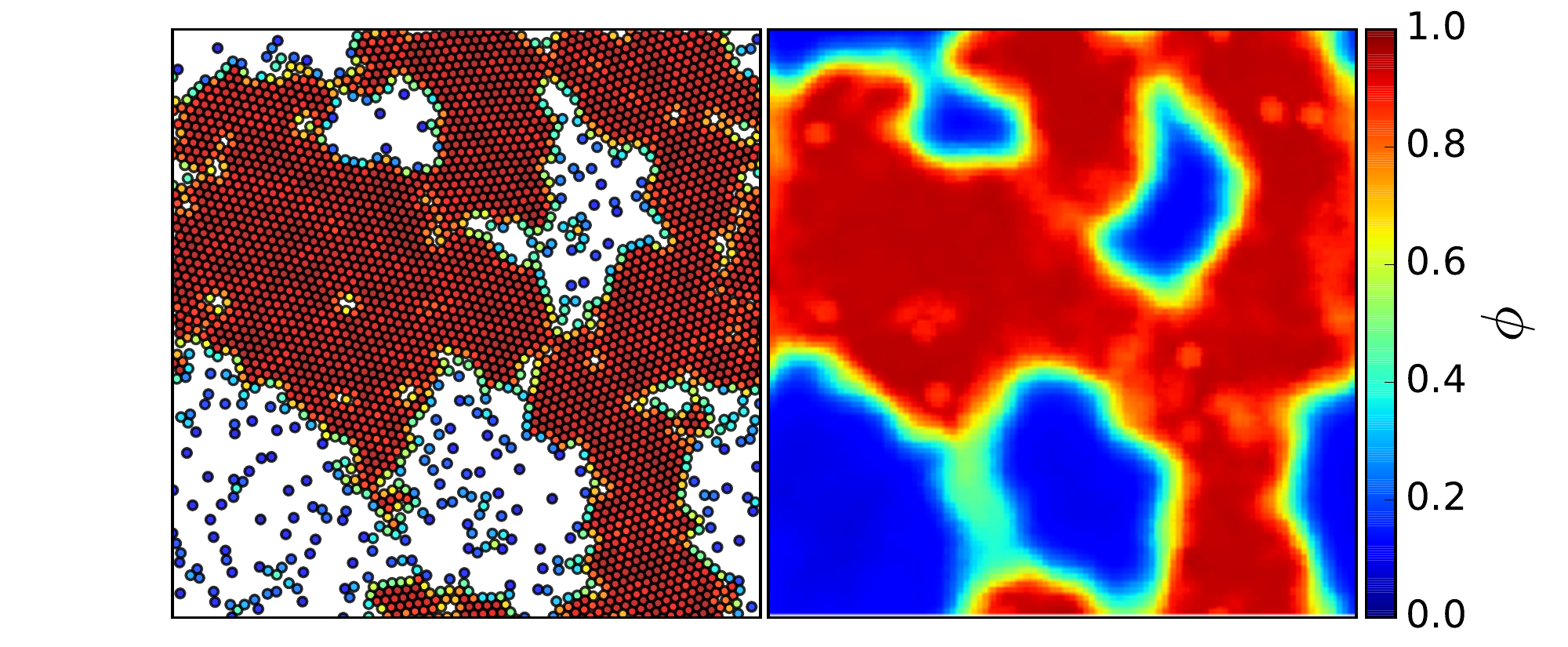}
	\includegraphics[width=0.45\textwidth]{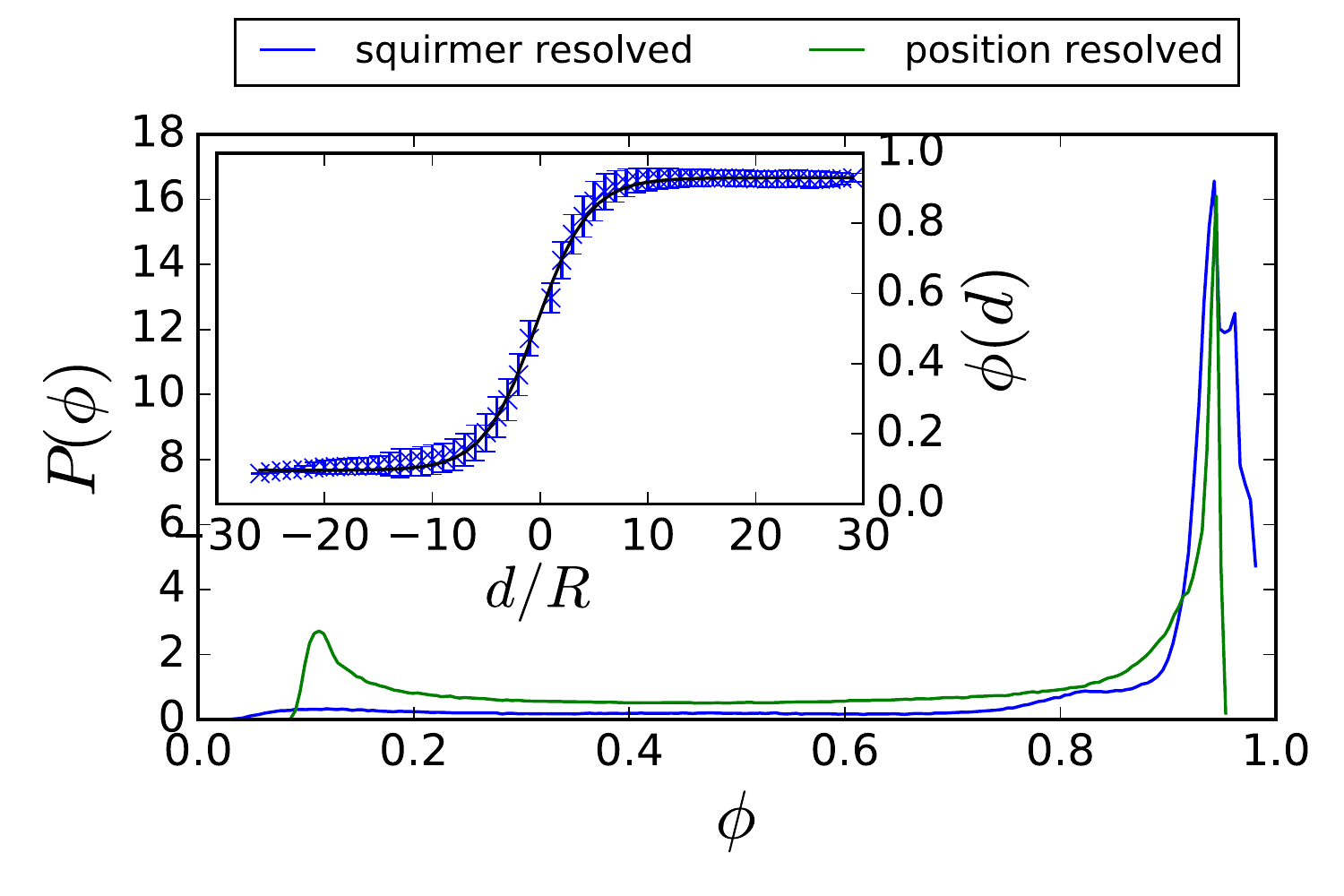}
	\caption{\emph{Top, left}: Snapshot of squirmer configuration for 3328 squirmers with $\mathrm{Pe} = 355$ ($B_1 = 0.1$) and $\bar{\phi} = 0.64$ exhibiting coexistence between a high-density and a low-density phase. 
	The colour code represents the local density $\phi_i$ around the $i$--th squirmer. 
	\emph{Top, right}: Corresponding position-resolved density $\phi(x,y)$ averaged over $3\times 10^6$ MPCD time steps. 
	\emph{Bottom, main plot}: Probability distribution of local density $\phi_i$ (blue line). 
	For comparison, the probability of the position-resolved density $\phi(x,y)$ is shown (green line).
	\emph{Inset}: Local density $\phi(d)$ as a function of distance from an interface, $d$, measured in units of $R$. 
	The solid black line is a fit to the interfacial density predicted by Cahn-Hilliard theory.
	\label{fig:a}
}
\end{figure}

For a range of total densities and P\'eclet numbers, the system decomposes into a dilute gas and a dense phase.
Figure\ \ref{fig:a}: \emph{top, left} shows a typical snapshot of a phase-separated system with mean area fraction $\bar{\phi} = 0.64$ and at P\'eclet number $\mathrm{Pe} = 355$ ($B_1=0.1$).
The two phases are clearly visible: many squirmers have come together to form a densely packed and system-spanning cluster, leaving a few squirmers 
in a much less dense ``gas'' phase. 
For lower $\bar{\phi}$, the dense phase consists of fewer squirmers and thus might not necessarily form system-spanning clusters. 
At even lower $\bar{\phi}$ phase separation does not occur and only the pure gas phase is visible.
The supplemental material$^\dag$ contains videos M1 and M2b showing a non phase-separating and phase-separating squirmer system, respectively. As well as the onset of phase separation M2a.

To quantify the local density, we construct around each squirmer a Voronoi cell and determine its area $A_i$.
The local density within this Voronoi cell is therefore $\phi_i = 1/A_i$. 
A standard analysis, such as the one used in \cite{Redner:2013jo}, would be to determine the probability distribution $p(\phi_i)$ of local densities $\phi_i$ and to identify the densities of the two coexisting phases from the expected bimodal distribution. 
The probability distribution corresponding to the squirmer snapshot in Fig.\ \ref{fig:a} is shown in Fig.\ \ref{fig:a}: \emph{bottom} by the blue line. 
The dense phase produces a very pronounced peak with a very broad tail, which obscures the signature of the gas phase.  
This broad tail is due to encounters between a small number of swimmers in the gas phase forming small temporary clusters.
Thus, some form of temporal averaging is needed, in order to keep only those dense regions that remain intact and relatively fixed in place. 

Hence we divide the system area into a grid with cell spacing equal to the squirmer radius $R$. 
Each cell is assigned a local density using a weighted average of the local densities from those Voronoi cells, which intersect each grid cell.
We then average over time, choosing the averaging interval to be longer than the life time of temporary clusters, yet short enough so that permanent clusters seem relatively unchanged.
Hence, we arrive at a position-resolved density $\phi(x,y)$ such as the one plotted in Fig.\ \ref{fig:a}: \emph{top, right}.
A histogram of the values of $\phi(x,y)$ is plotted in Fig.\ \ref{fig:a}: \emph{bottom},green line, and recovers the peak for the gas phase.

\begin{figure}
	\centering
	\includegraphics[width=0.4\textwidth]{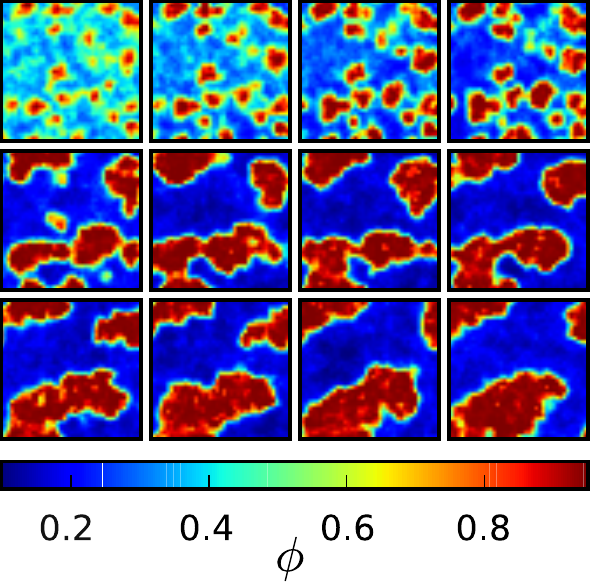}
	\includegraphics[width=0.48\textwidth]{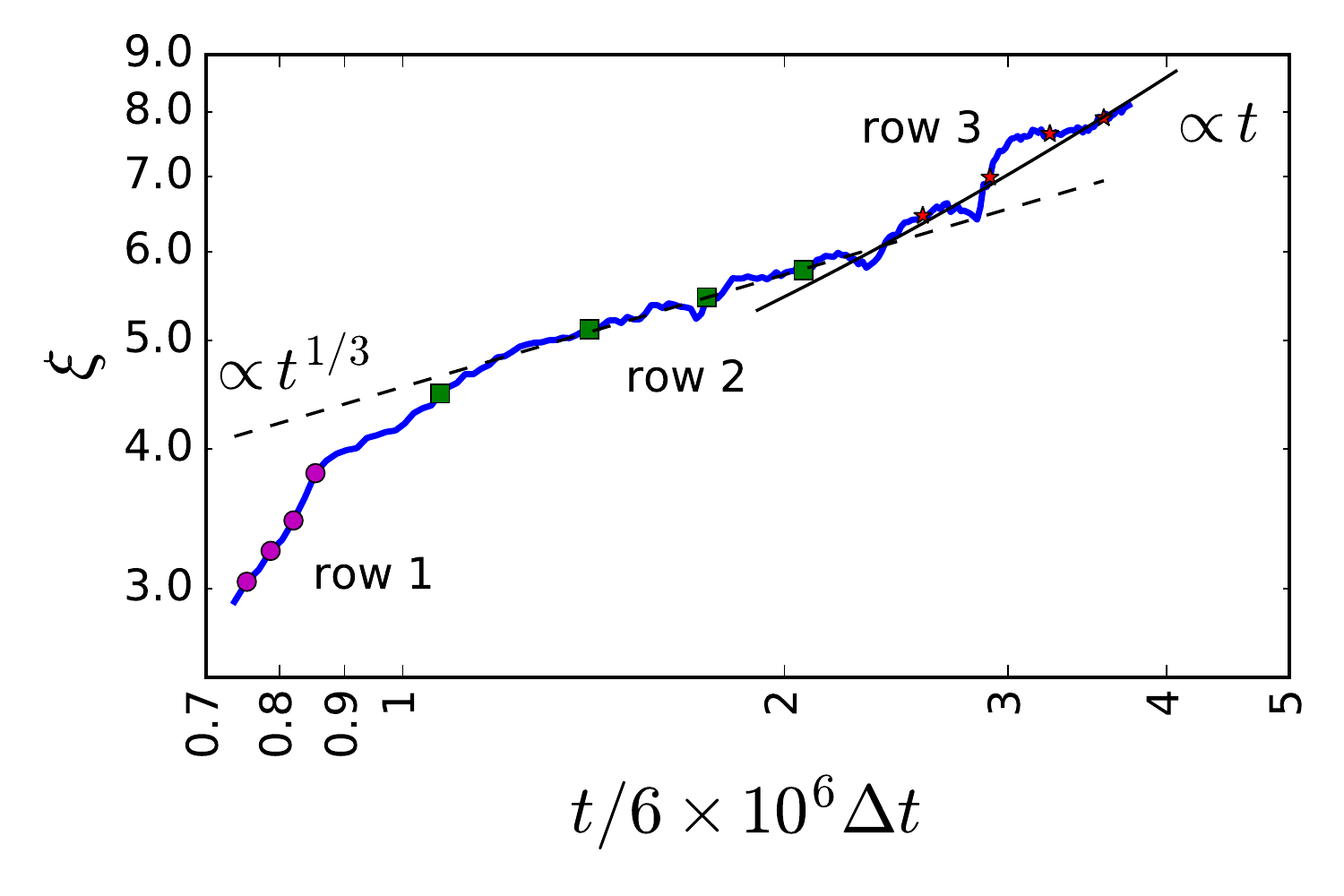}
	\caption{\emph{Top:} Time evolution of $\phi(x,y)$.
	Each successive frame represents a time marked on the plot of $\xi(t)$ (bottom panel). 
	The first, second, and third row are marked by magenta circles, green squares, and red stars, respectively.
	 \emph{Bottom:} Characteristic cluster size $\xi(t)$ as a function of time (in units of MPCD time step $\Delta t$). 
	 Simulation data for $\bar{\phi} = 0.48$ is shown in the blue solid line. 
	 The coarsening dynamics of model H for diffusive transport, $\xi(t)\sim t^{1/3}$, and viscous hydrodynamic transport, $\xi(t)\sim t $, are indicated by the dashed and solid black lines, respectively. \label{fig:phi-time_series}}
\end{figure}

The color-coded density $\phi(x,y)$ in Fig.\ \ref{fig:a} also shows a pronounced interfacial region between dense and gas phase.
In order to obtain their densities and distinguish them from the interface, we determine the interfacial density profile $\phi(d)$, where $|d|$ is the distance to the nearest point on the interface. 
A preliminary position of the interface is determined by requiring that the time-averaged number of neighbours is three. 
Variations in this choice simply shift the density profile along the $d$-axis, but do not alter its shape.
Thus the resulting $\phi(d)$ is independent of fixing the exact interface position beforehand.
For each grid cell the distance to each point of the estimated interface is determined and the minimal value is chosen as $|d|$. 
Collecting the statistics for each grid cell then gives $\phi(d)$.
The resulting profile corresponding to the color-coded density $\phi(x,y)$ can be seen in the inset of Fig.\ \ref{fig:a}: \emph{bottom}.
Inspired by Ref. \cite{Bialke:2015cq} and Cahn-Hilliard theory \cite{Wittkowski:2014dt}, we fit 
\begin{equation}
	\phi(d) = \frac{\phi_\mathrm{den}+ \phi_\mathrm{gas}}{2} + \frac{\phi_\mathrm{den} - \phi_\mathrm{gas}}{2}\tanh{\left(\frac{d-d_0}{2w}\right)}
	\label{eq:phi-d}
\end{equation}
to the profile in the inset (\emph{cf.} solid line).
This provides values for the respective densities of gas and dense phase, $\phi_\mathrm{gas}$ and $\phi_\mathrm{den}$.
The good quality of the fit is not surprising, as an effective Cahn-Hilliard equation was formulated for the local density in systems comprised of active Brownian particles \cite{Speck:2014iy,Speck:2015gc}.

Before we introduce the resulting phase diagram, we discuss the coarsening dynamics from the uniform initial state to the final 
phase-separated system. 
Figure\ \ref{fig:phi-time_series}: \emph{top} illustrates the time evolution for a system with $\bar{\phi} = 0.48$ and $\mathrm{Pe} = 355$ ($B_1=0.1$) on different time scales.
Initially (first row) clusters of squirmers develop in the unstable uniform phase characteristic for spinodal decomposition.
They grow and become denser until a few clusters remain (second row). These clusters already have the final density 
$\phi_\mathrm{den}$ of the dense phase. 
They show the typical coarsening dynamics of a phase-separating system due to diffusive exchange of squirmers between the clusters. 
Ultimately, one single cluster remains (third row), the shape of which becomes more compact over the course of time.

Following Ref.\ \cite{Bray:2003ek}, we quantify coarsening by determining the area $a(t)$ of all dense regions and their total perimeter $l(t)$ as a function of time. 
The mean domain size then becomes $\xi(t) := a(t) / l(t)$. Figure\ \ref{fig:phi-time_series}: \emph{bottom} compares $\xi(t)$ to the different regimes of the coarsening dynamics for a system with conserved order parameter including hydrodynamics\ \cite{Bray:2003ek}. 
After the initial formation of the clusters (the magenta dots refer to the frames in the first row), the total area $a(t)$ remains constant and all further changes in $\xi$ are due to changes in $l(t)$ alone.
In the intermediate regime, the mean domain size roughly obeys the scaling expected when diffusive transport dominates, $\xi\sim t^{1/3}$ (frames in the second row). 
This regime lasts less than a decade, since our system size is relatively small.
Former work on the coarsening dynamics of pure active Brownian particles report a deviation from the scaling exponent $1/3$, measuring 0.255 \cite{Redner:2013jo} or 0.28 \cite{Stenhammar:2013kb}.
However, our system is too small to make a clear statement about the exponent.
Finally, the compactification of the single cluster gives rise to clearly visible, jumplike increases of $\xi(t)$. 
For comparison, we give the expected scaling $\xi\sim t$ due to viscous hydrodynamic transport.
This compactification has not been seen for pure active Brownian particles.
Also, in our simulations not all $\bar{\phi}$ enter this regime. 
Small clusters leave the diffusive transport regime with fairly circular clusters already. 
Whereas for $\bar{\phi}\geq 0.50$ clusters spanning the whole simulation box are formed. 
These are stable, either as ``slabs'' for $\bar{\phi} \approx 0.50$ or the gas phase forms a ``bubble'' at even larger mean densities.
The rapid increase of $\xi(t)$, shown in Fig.~\ref{fig:phi-time_series}: \emph{bottom} towards the end, therefore requires $\bar{\phi}\lesssim 0.50$ and probably is a finite size effect.

\subsection{Phase diagram of neutral squirmers}

\begin{figure}
	\includegraphics[width=0.49\textwidth]{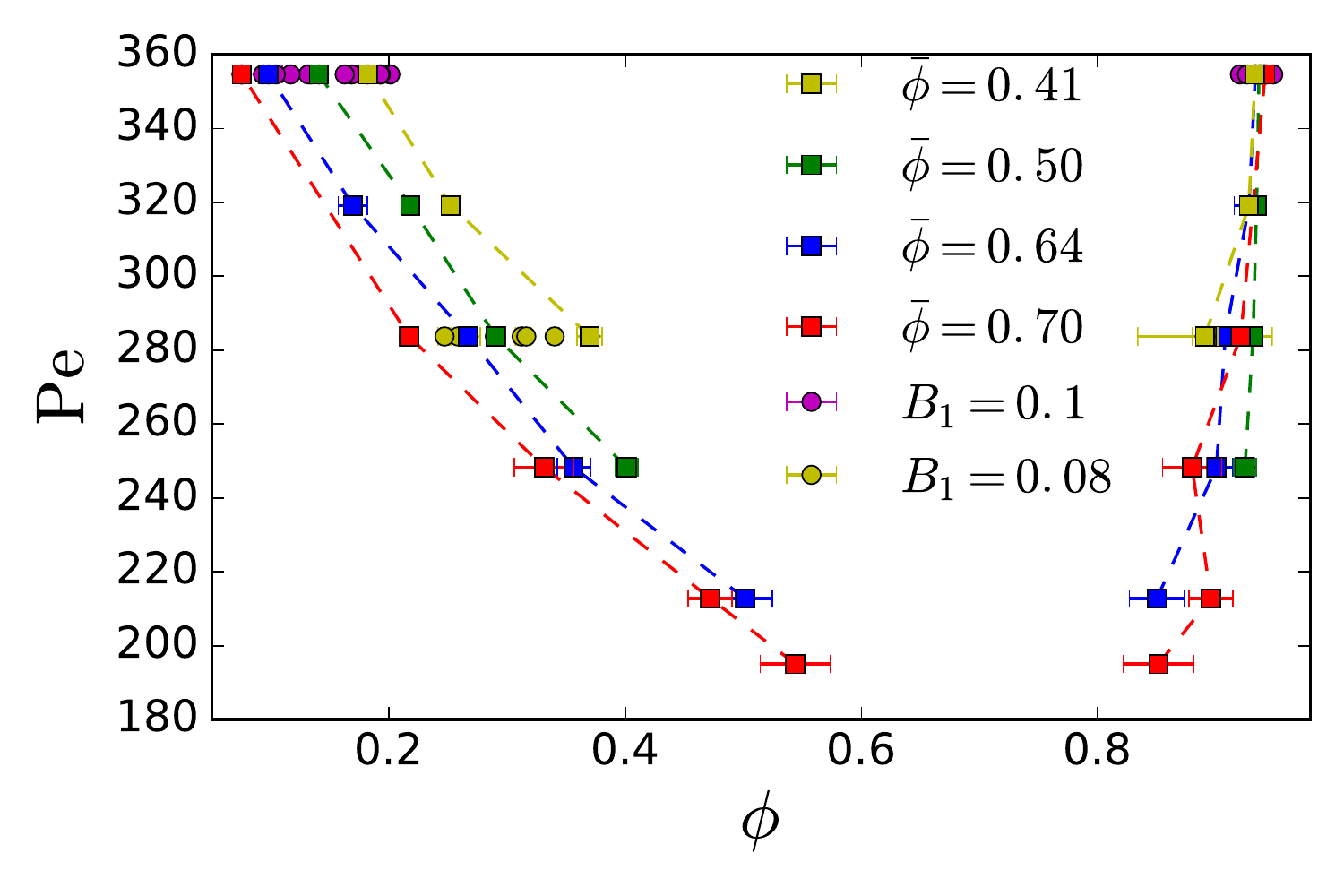}
	\includegraphics[width=0.49\textwidth]{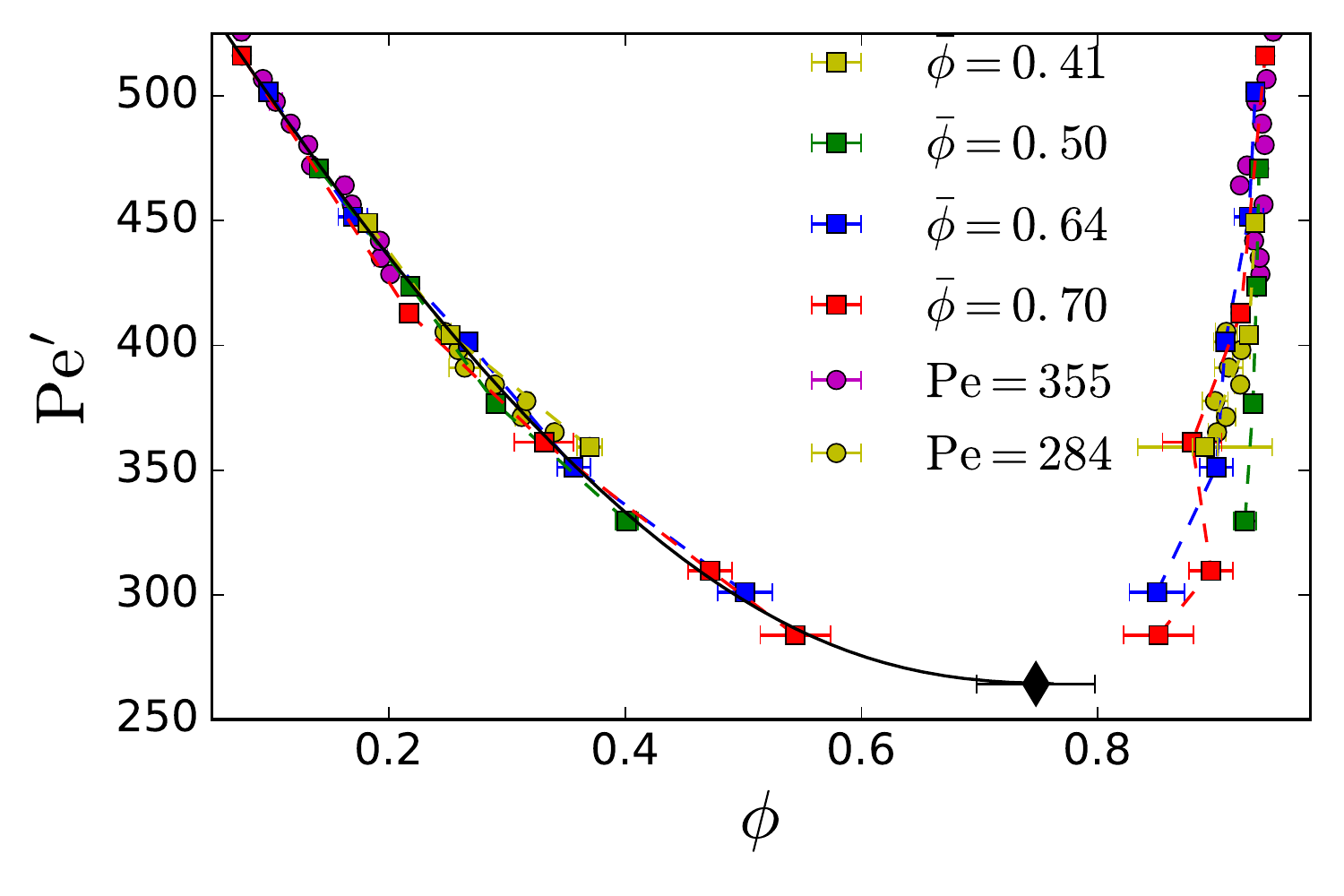}
	\caption{\emph{Top:} Binodal lines showing the coexistence densities of neutral squirmers for different mean densities $\bar{\phi}$.
	The magenta points for $\mathrm{Pe} = 355$ correspond to simulations from Fig.\ \ref{fig:phi-gas}. 
         \emph{Bottom:} The rescaled data using the effective P\'eclet number $\mathrm{Pe}^\prime = \left( 1 + 0.65 \bar{\phi}\right) \mathrm{Pe}$ shows a data collapse for different $\bar{\phi}$. 
         The black line shows a fit to Eq.~\eqref{eq:fit_gas}. 
         The fitted critical point is marked by the back diamond. \label{fig:coex}
	}
\end{figure}

\begin{figure}
	\includegraphics[width=0.5\textwidth]{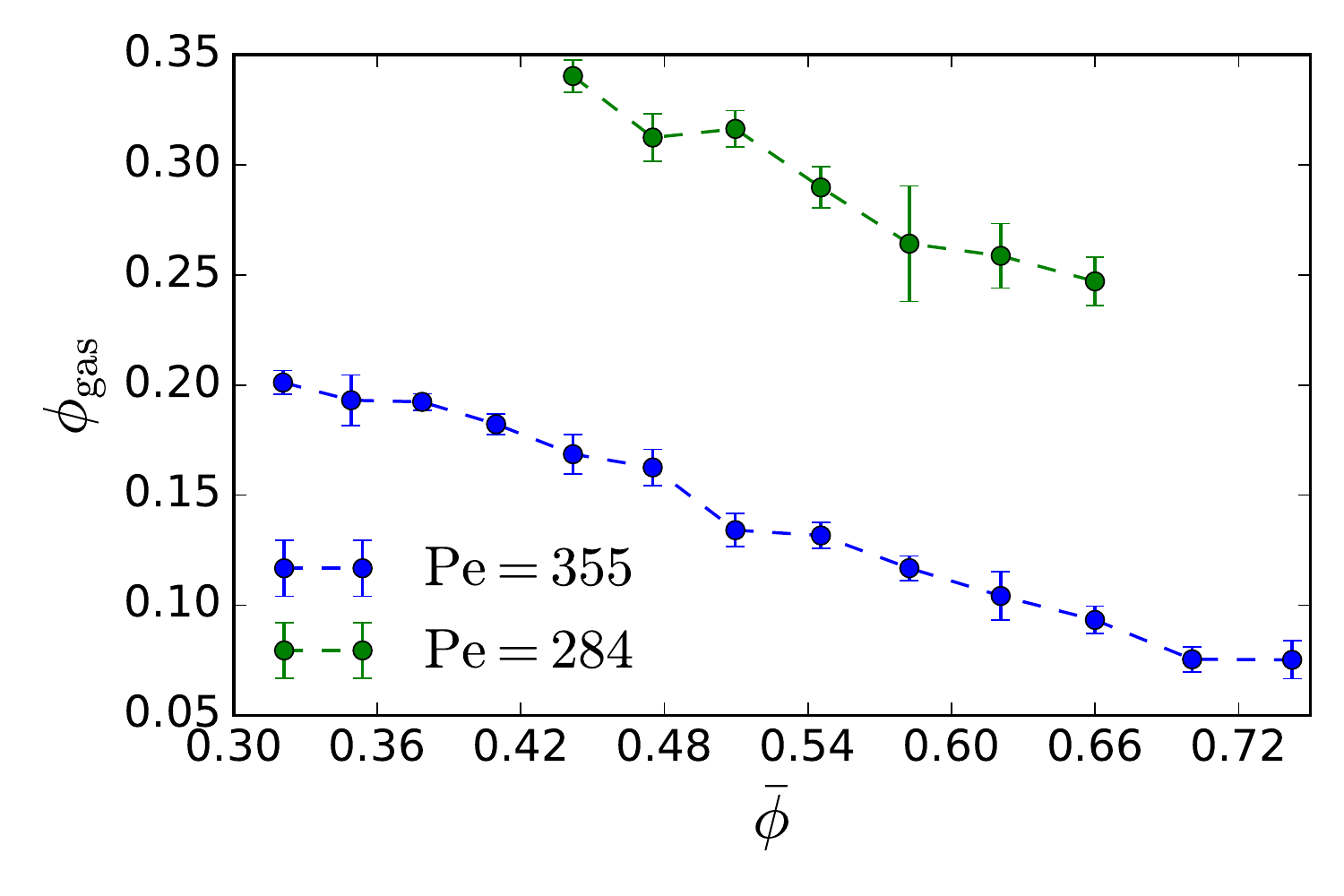}
	\caption{Gas density $\phi_{\mathrm{gas}}$ in the phase-separated system plotted versus mean density $\bar{\phi}$ for neutral 
	squirmers with $\mathrm{Pe} = 355$ ($B_1=0.1$) and $\mathrm{Pe} = 284$ ($B_1=0.08$).
	\label{fig:phi-gas}}
\end{figure}

In Fig.\ \ref{fig:coex}:\ \emph{top} we plot the densities $\phi_\mathrm{gas}$ and $\phi_\mathrm{den}$ of the coexisting gas and dense phases in a $\mathrm{Pe}-\phi$ diagram, which we determined for different mean densities $\bar{\phi}$. 
Interestingly, the resulting binodal lines do not agree. 
In particular, the gas binodal is shifted to smaller values $\phi_\mathrm{gas}$ for increasing $\bar{\phi}$.
In contrast, for pure active Brownian particles, the densities of the two coexisting phases are independent of $\bar{\phi}$ \cite{Redner:2013jo,Speck:2015gc}. 
For our geometry we present them in appendix\ \ref{sec.comp}.
Thus, we suspect hydrodynamic interactions between the squirmers to cause this behaviour.
For constant P\'eclet number Fig.\ \ref{fig:phi-gas} demonstrates how the gas density $\phi_{\mathrm{gas}}$ decreases for larger $\bar{\phi}$. 
Thus, the cluster of dense phase, which grows in size with increasing $\bar{\phi}$, extracts additional squirmers from the gas phase. 
Below we will attribute this behaviour to an additional hydrodynamic pressure caused by the squirmers at the interface between
the dense and the dilute phase.

The gas binodals in Fig.\ \ref{fig:coex}:\ \emph{top} roughly run parallel to each other. 
Thus, by introducing an effective P\'eclet number
\begin{equation}
	\mathrm{Pe}^\prime = \left( 1 + a \bar{\phi}\right) \mathrm{Pe}  \label{eq:B1:scaling}
\end{equation}
with fit parameter $a \approx 0.65$, we are able to collapse them on a single master curve for the binodal line as Fig.\ \ref{fig:coex}: \emph{bottom} demonstrates.
It is tempting to explain the coefficient $a$ by defining an effective P\'eclet number using a density-dependent diffusion coefficient $D_{T,2D} (\bar{\phi}) = D_{T,2D} (\bar{\phi} =0) / (1+2.027\bar{\phi})$
in the definition (\ref{eq.Pe}) of the P\'eclet number. We determined it from the velocity-auto-correlation
function and the Green-Kubo formula (see also paragraph after Eq.\ (\ref{eq:p_act}) and Appendix\ \ref{app.pressure}). 
However, the prediction of 2.027 for the coefficient $a$ cannot explain the reported data collapse.

In order to estimate the position of the critical point $(\phi_*, \mathrm{Pe}^\prime_*)$, we fit \cite{Speck:2015gc}
\begin{equation}
	\frac{\phi_\mathrm{gas}}{\phi_*} = 1 + a\left( \frac{\mathrm{Pe}^\prime}{\mathrm{Pe}^\prime_*} \right) + b\left( \frac{\mathrm{Pe}^\prime}{\mathrm{Pe}^\prime_*} \right)^\frac{1}{3} \;, \label{eq:fit_gas}
\end{equation}
to the gas-binodal in Fig.~\ref{fig:coex}:\ \emph{bottom}.
The supplemental material$^\dag$ contains videos M3 and M4 of the dynamics of the position-resolved density $\phi(x,y)$ inside the phase-coexistence regime and outside of it close to the critical point.

For a system in mechanical equilibrium the total pressure in the coexisting gas and dense phases has to be the same.
We now apply this strategy to our system of phase-separating squirmers in order to understand why the binodals depend on the mean density.
For active Brownian particles without any hydrodynamic interactions, the pressure balance was first proposed in Ref.\ \cite{Takatori:2014do}
by introducing a total pressure consisting of a steric and an active contribution: $p = p^\mathrm{(s)}+p^\mathrm{(a)}$.
This pressure balance was then used in the context of phase separation in Refs.\ \cite{Takatori:2014do,Bialke:2015cq,Solon:2015hz,Solon:2015bt,Winkler:2015cy,Takatori:2015bp}.
The steric pressure results from direct interactions between the squirmers, $p^\mathrm{(s)} = \langle \sum_{ij} \vv{r}_{ij} \cdot \vv{F}_{ij}\rangle / (6V)$ \cite{Tsai:1979dj}.
Here,  $\vv{r}_{ij} = \vv{r}_{i} - \vv{r}_{j}$ is the distance vector from swimmer $j$ to swimmer $i$, $\vv{F}_{ij}$ the force with which squirmer $j$ acts on $i$, $V$ is the volume of a uniform system, and $\langle ... \rangle$ means temporal average. 
Swimmers moving with constant velocity $v_0$ along the unit vector $\hat{\vv{e}}_i$ also generate a swim pressure\cite{Takatori:2014do}:
\begin{equation}
	p^\mathrm{(a)} = \frac{v_0\gamma_0}{3 V} \sum_{i=1}^N  \langle \hat{\vv{e}}_i \cdot \vv{r}_i   \rangle \;,    \label{eq:p_act}
\end{equation}
where $\gamma_0$ is a friction coefficient connected to energy dissipated by the swimmer in its fluid environment.
We note here that the swim pressure is only defined over times, which are long compared to $1/D_\mathrm{R}$, therefore the average $\langle\cdots\rangle$ needs to be taken over sufficiently long times\cite{Brady16}.
%
%
As swimmers move, while this averaging is taking place, $p^\mathrm{(a)}$ becomes ``smeared out'' and therefore describes the active pressure in the volume $V$ and not at a point.
\footnote{One can argue that the linear exention of the volume $V$ should be larger than the persistence length of the swimmer.
However, if one averages over a sufficiently long time so that a sufficient number of swimmers enter and leave $V$, this requirment might be relaxed. 
This needs to be checked more carefully in a future investigation.}

Due to hydrodynamic interactions with bounding walls and other squirmers,
the friction coefficient in Eq.~\eqref{eq:p_act} depends not only on the squirmer's position but also the position of every other squirmer.
Hence, it is not possible to know $\gamma_0$ precisely. 
To arrive at a sort of mean-field approximation for $\gamma_0(\phi)$, we use the Green-Kubo formula and determine a self-diffusion coefficient $D_{T,2D} (\phi)$ from the 
velocity-auto-correlation function of a passive particle with radius $R$ using the velocity components in the plane of the 
quasi-two-dimensional geometry. The result is plotted in Fig.\ \ref{fig.selfdiff} of 
Appendix\ \ref{app.pressure}
and can be well approximated by $D_{T,2D} (\phi) = D_{T,2D} (\phi =0) / (1+2.027\phi)$. We then use the Einstein relation, $\gamma_0(\phi) = k_\mathrm{B}T /D_{\mathrm{T}, 2D}(\phi)$, to arrive at a density-dependent friction coefficient. 
However, the following results only change little compared to working with a constant $\gamma_0 = k_\mathrm{B}T /D_{\mathrm{T}, 2D}(\phi=0)$ for all densities.

In a dilute gas of non-interacting Brownian particles, orientation decorrelates exponentially on the 
decorrelation time $2 \tau_r$ and one obtains from Eq.\ (\ref{eq:p_act}) \cite{Takatori:2014do}
\begin{equation}
p^\mathrm{(a)}_\mathrm{id} = \rho_\mathrm{gas} \tau_r \gamma_0 v_0^2 / 6  \sim \rho_\mathrm{gas} \mathrm{Pe}^2 \, ,
		\label{eq:p_act_id}
\end{equation}
where $\rho_\mathrm{gas}$ is number density. 
This active gas pressure is controlled by the P\'eclet number. 
It holds the dense phase of active Brownian particles together by preventing them from swimming apart.

\begin{figure}
	\includegraphics[width=0.5\textwidth]{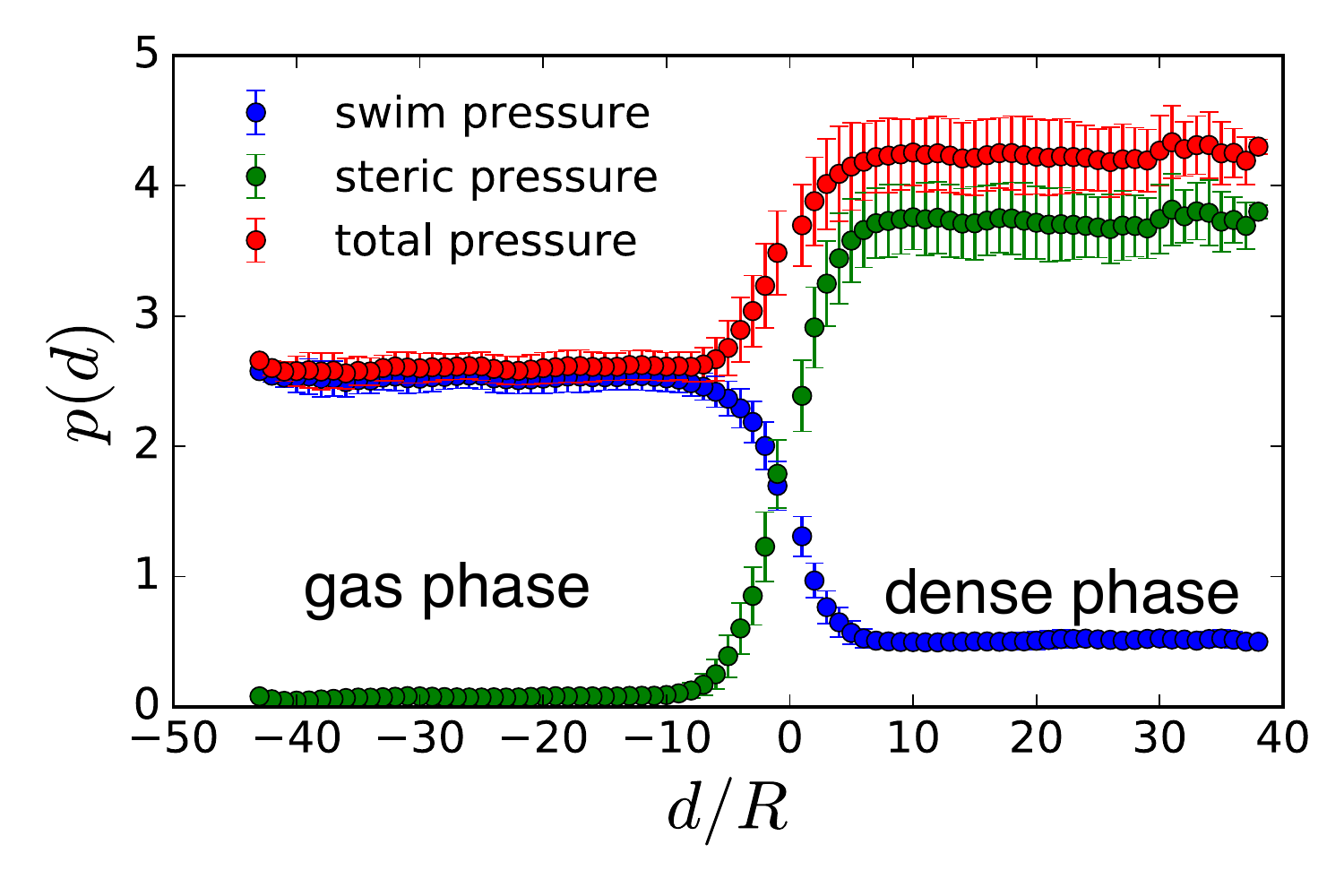}
	\includegraphics[width=0.5\textwidth]{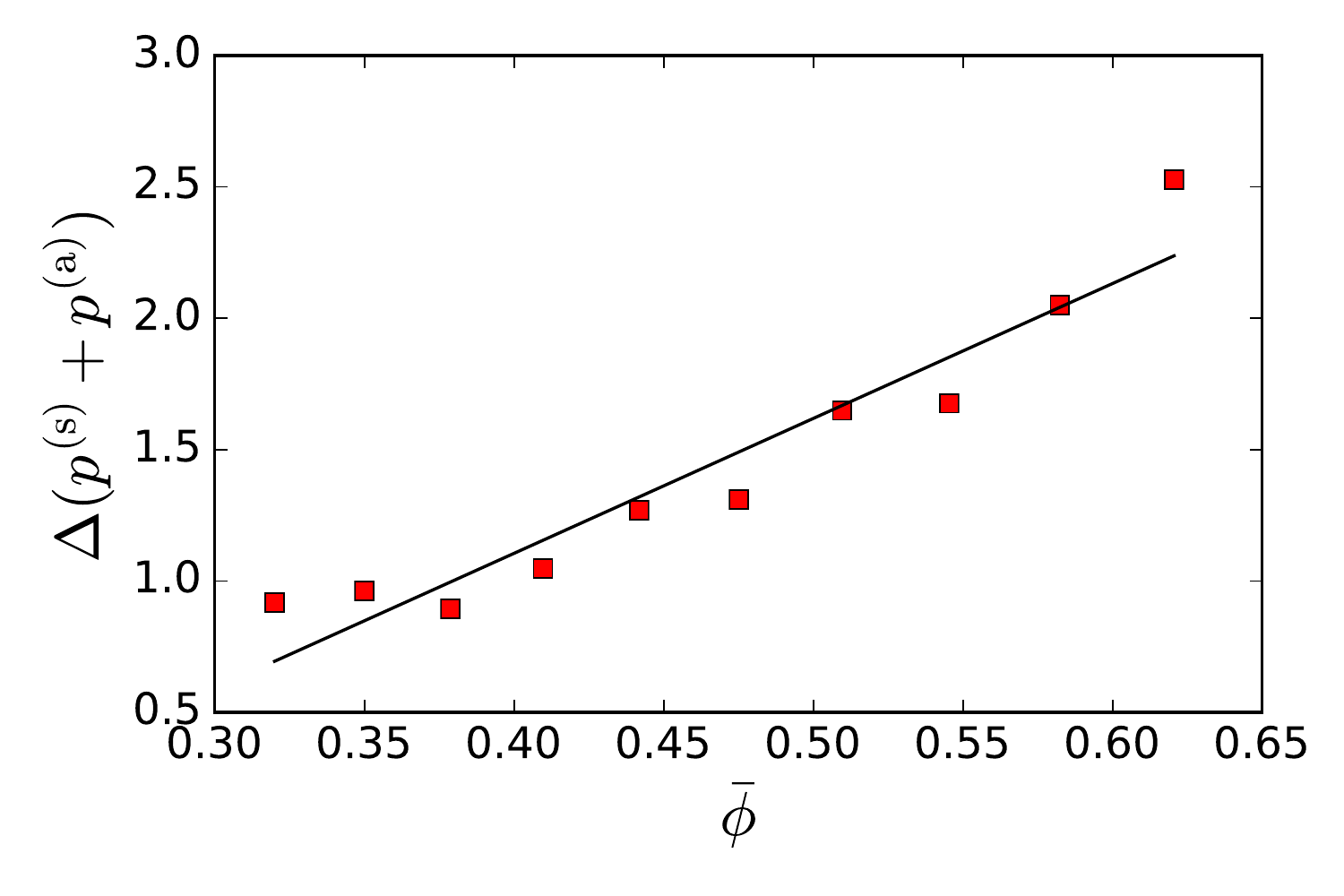}
	\caption{\emph{Top:} Local pressure values plotted across the interface between gas and dense phase for $\mathrm{Pe}=355$ ($B_1=0.1$) and $\bar{\phi}=0.51$. 
	Distance from the interface $d$ is measured in units of squirmer radius $R$.
	$d>0$ represents the interior of the cluster formed by the dense phase.
	\emph{Bottom:} Total pressure jump $\Delta (p^\mathrm{(s)} + p^\mathrm{(a)})$ when going into the dense phase plotted versus $\bar{\phi}$.
	The black line shows the fit $\Delta p(\bar{\phi}) = 5.1\bar{\phi} - 0.95$. \label{fig:pressure}}
\end{figure}

We are able to calculate steric and the swim pressure of Eq.~\eqref{eq:p_act} for the local neighbourhood of each squirmer (see Appendix\ in\ section~\ref{sec:appendix} for details).
Figure\ \ref{fig:pressure}:\ \emph{top} shows these local pressure values as a function of the distance from the gas-cluster interface. 
In light of the discussion about the swim pressure following Eq.\ (\ref{eq:p_act}), its pressure values close to the interface have to be treated with caution. 
But they are not important for the following.
Outside the dense phase, $d<0$, the steric pressure is low due to very few contacts between squirmers or a squirmer and a bounding wall.
And when they do occur, they persist for only a short time.
On the other hand, the local swim pressure in the gas phase is large as swimmers can move unhindered.
Inside the cluster, $d>0$, squirmers are in constant contact with their neighbours and the confining walls.
This leads to a significantly higher steric pressure. 
On the other hand, the squirmer's swimming motion is hindered by the neighbours.
Swimming motion and actual displacement are much less correlated and the swim pressure is greatly reduced.

\begin{figure}
	\includegraphics[width=0.49\textwidth]{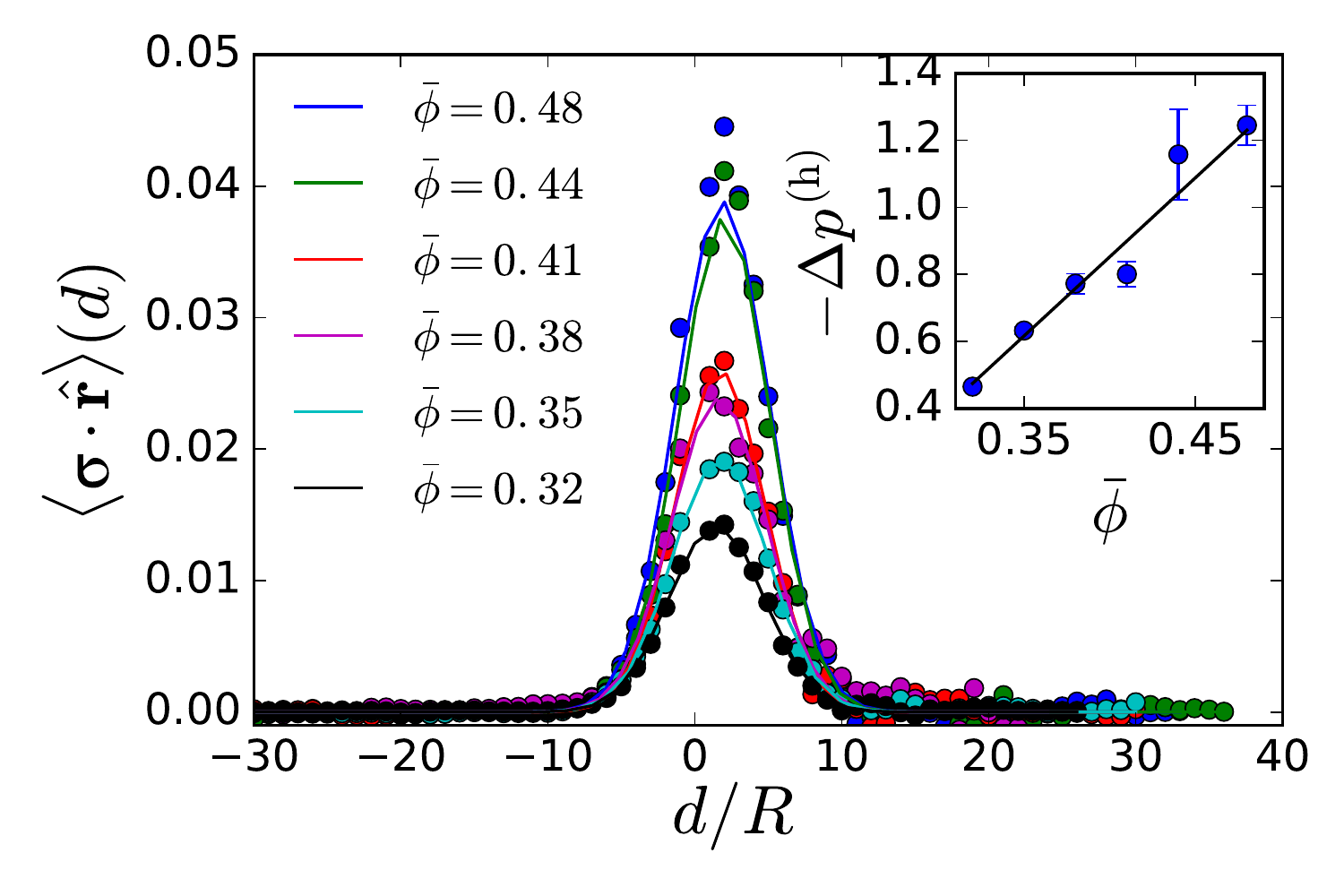}
	\caption{\emph{Top:} Radial source-dipole density plotted versus distance from the cluster interface, $d$, measured in units of squirmer radius $R$ for different mean densities $\bar{\phi}$. 
	\emph{Inset:} Magnitude of resulting hydrodynamic pressure difference versus $\bar{\phi}$.
	The solid line shows a fit $-\Delta p^\mathrm{(h)} = 4.72\bar{\phi} - 1.04$.
	\label{fig:orientation}
	}
\end{figure}

However, contrary to what is expected for mechanical equilibrium, the total pressure $p = p^\mathrm{(s)}+p^\mathrm{(a)}$ plotted in Fig.\ \ref{fig:pressure}:\ \emph{top} is not the same for the gas and the dense phase. 
Furthermore, in Fig.\ \ref{fig:pressure}:\ \emph{bottom} we see that this total pressure difference $\Delta p = p_\mathrm{den}-p_\mathrm{gas}$ across the phase interface increases with increasing $\bar{\phi}$. 
Thus, there must be an additional negative pressure stabilizing this interface.
When introducing the swim pressure, the authors of \cite{Takatori:2014do} already remarked that for swimmers swimming in a fluid environment, the hydrodynamic stresslet induces an additional hydrodynamic pressure.
In the following we elaborate on this idea for our special geometry and demonstrate that indeed we are able to justify a negative pressure difference between the dense and the gas phase.

It is not possible to give a complete quantitative theory for the negative hydrodynamic pressure produced by the dense cluster of squirmers.
Instead, we present here an approximate treatment, which contains the essential ideas. 
A squirmer at position $\vv{r}_0$ and confined between two plates generates the flow field of a source dipole with the approximate dipole moment $\bm{\sigma} = 2\pi R^2 v_0 \hat{\vv{e}}$ \cite{Brotto:2013fj,Delfau:2016cv}.
It is accompanied by the pressure field
\begin{equation}
	p(\vv{r}) = - \frac{1}{2\pi G} \frac{\bm{\sigma} \cdot (\vv{r} - \vv{r}_0)}{| \vv{r} - \vv{r}_0  |^2  }  \, , \label{eq:p_2D}
\end{equation}
where $G\sim H^2 / \eta$ measures the system's confinement with $H$ the plate distance and $\eta$ viscosity.
To be concrete, we use the value for a Poiseuille flow in MPCD units, $G=H^2 /12 \eta \approx 1/3$.

Now, we consider a circular cluster of squirmers and replace the dipole moment $\bm{\sigma}$ by a density of source dipoles.
Furthermore, we find that on average the squirmers have a tendency to point radially inwards towards the center, especially at the rim of a cluster. 
Therefore, we introduce a source dipole density $\bar{\bm{\sigma}} = \langle  \bm{\sigma} \cdot \hat{\vv{r}}_c \rangle \hat{\vv{r}}_c$, where $\hat{\vv{r}}_c$ is a unit vector pointing from the squirmer to the cluster center. 
The radial dipole density $\langle  \bm{\sigma} \cdot \hat{\vv{r}}_c \rangle $ across the interface from the gas to the dense phase is plotted in Fig.\ \ref{fig:orientation} for different $\bar{\phi}$.
This confirms the preferred radial alignment at the rim of the dense cluster, whereas in the interior the squirmers point towards the bounding walls, on average, since $\langle  \bm{\sigma} \cdot \hat{\vv{r}}_c \rangle $ is zero.
We calculate the pressure generated by the dipole density in two steps. 
First, we integrate along a circle of radius $r_0$ to obtain pressure per unit length:
\begin{equation}
p(\vv{r}; r_0) = - \frac{1}{2\pi G}		
\oint  \frac{\bar{\bm{\sigma}} \cdot (\vv{r} - \vv{r}_0)}{| \vv{r} - \vv{r}_0|^2 } \dd s \,
\end{equation}
With $\vv{r}_0 = (r_0 \cos{\varphi}, r_0\sin{\varphi})$, $\cos \varphi = \vv{r}_0 \cdot \vv{r}$, and $\vv{r}_0 = - r_0 \hat{\vv{r}}_c$ one obtains
\begin{equation}
p(\vv{r}; r_0)  = -\frac{\langle \bm{\sigma} \cdot \hat{\vv{r}}_c \rangle(r_0)}{2\pi G} \int_0^{2\pi}  \frac{r_0 - r \cos{\varphi}}{{r^2+r_0^2
 - 2 r r_0\cos{\varphi}}} r_0 \dd{\varphi}  \, ,
\end{equation}
which can be integrated to
\begin{equation}
p(\vv{r}; r_0) = -\frac{\langle \bm{\sigma}\cdot\hat{\vv{r}}_c \rangle(r_0)}{2\pi G} \left\{ 
		\begin{array}{cc}
			2\pi & r<r_0\\
			0 & r>r_0
		\end{array} 
		\right.\, .
	\label{eq:ph_deriv}%
\end{equation}
Thus inside a ring of source dipoles pointing radially inwards, pressure is constant and negative, while outside of the ring it is zero.
Second, we integrate over the radial coordinate and obtain the constant hydrodynamic pressure difference between cluster interior and exterior,
\begin{equation}
	\Delta p^\mathrm{(h)} = - \frac{1}{G}  \int_0^{R_c} \langle \bm{\sigma} \cdot \hat{\vv{r}}_c \rangle(r_0) \dd r_0 \, , \label{eq:dph}
\end{equation}
where $R_c$ is the cluster radius. 
This negative pressure should cancel the difference of the total pressure between dense and gas phases, which is illustrated in Fig.\ \ref{fig:pressure}: \emph{top}.

In summary, our argument is as follows: any net polar order of swimmers is accompanied by a jump in solvent pressure, a fact that has also been reported by Ref.\ \cite{Yan:2015fg}.
In our case, this is due to the squirmers at the edge of the cluster: the hydrodynamic pressure difference between cluster interior and exterior is necessary to stop the flow initiated by the squirmers, which try to pump fluid out of the cluster interior.

The inset in Fig.\ \ref{fig:orientation} shows the hydrodynamic pressure using the radial source-dipole density $\langle \bm{\sigma} \cdot \hat{\vv{r}}_c \rangle$ from the main plot. 
Its absolute value increases linearly with cluster size or $\bar{\phi}$, which results from the stronger radial alignment as the cluster size increases.
The linear increase of $ | \Delta p^\mathrm{(h)} | $ coincides with the linear increase of $\Delta(p^\mathrm{(s)}+p^\mathrm{(a)})$ plotted in Fig.~\ref{fig:pressure}:\ \emph{bottom}, however the slope is by a factor of $1.1$ too low. 
This is understandable since we approximate the squirmers in the dense cluster by point-like source dipoles with an approximate dipole moment. 
We only calculated $ \Delta p^\mathrm{(h)} $ up to $ \bar{\phi} = 0.48 $, where the dense clusters are nearly circular. 
Beyond this mean density, they become elongated and the radial alignment of the squirmers cannot clearly be defined.

Thus our claim is that by introducing the negative hydrodynamic pressure generated by squirmes swimming into the cluster at its rim, we are able to maintain mechanical equilibrium. 
Namely, by defining the total pressure $p = p^\mathrm{(s)}+p^\mathrm{(a)}+p^\mathrm{(h)}$ with the hydrodynamic pressure included, this total pressure should be constant across the interface of gas and dense phase.
Our observation that $| \Delta p^\mathrm{(h)}|$ increases with cluster size and thus with $\bar{\phi}$, explains the different binodals in Fig.\ \ref{fig:pressure} and why we can collapse all of them on a single master curve with the help of the effective P\'eclet number from  Eq.\ \eqref{eq:B1:scaling}. 
For an increased $| \Delta p^\mathrm{(h)}|$, a lower $\Delta p^\mathrm{(a)} \propto \phi_\mathrm{gas}$ is 
required to maintain the interface and thus $\phi_\mathrm{gas}$ is reduced.

For phase-separating systems, interface curvature leads to a Laplace pressure and can result in a shift of the coexistence densities \cite{Statt:2015jy}.
We point out that this effect cannot be responsible for the shifted binodals as well as the pressure jump in Fig.\ \ref{fig:pressure}:\ \emph{bottom}.
First, the Laplace pressure changes signs as one goes from a droplet to a bubble configuration.
In particular, the pressure jump in Fig.\ \ref{fig:pressure}:\ \emph{top} was determined for the "slab" configuration with nearly straight interfaces, where the Laplace pressure should vanish. 
Second, the coexistence densities of the dilute and dense phase in the middle row of Fig.~\ref{fig:phi-time_series}:\ \emph{top} do not change while coarsening.
However, if Laplace pressure were important in our system, the densities would change since the curvature of the interfaces changes.

Until now we have only examined neutral squirmers. 
In the following we present binodal lines for different squirmer parameter $\beta$ to demonstrate how pushers and pullers effect the phase-coexistence region of microswimmers.

\subsection{Binodal Lines of Pushers and Pullers}

\begin{figure}
	\includegraphics[width=0.5\textwidth]{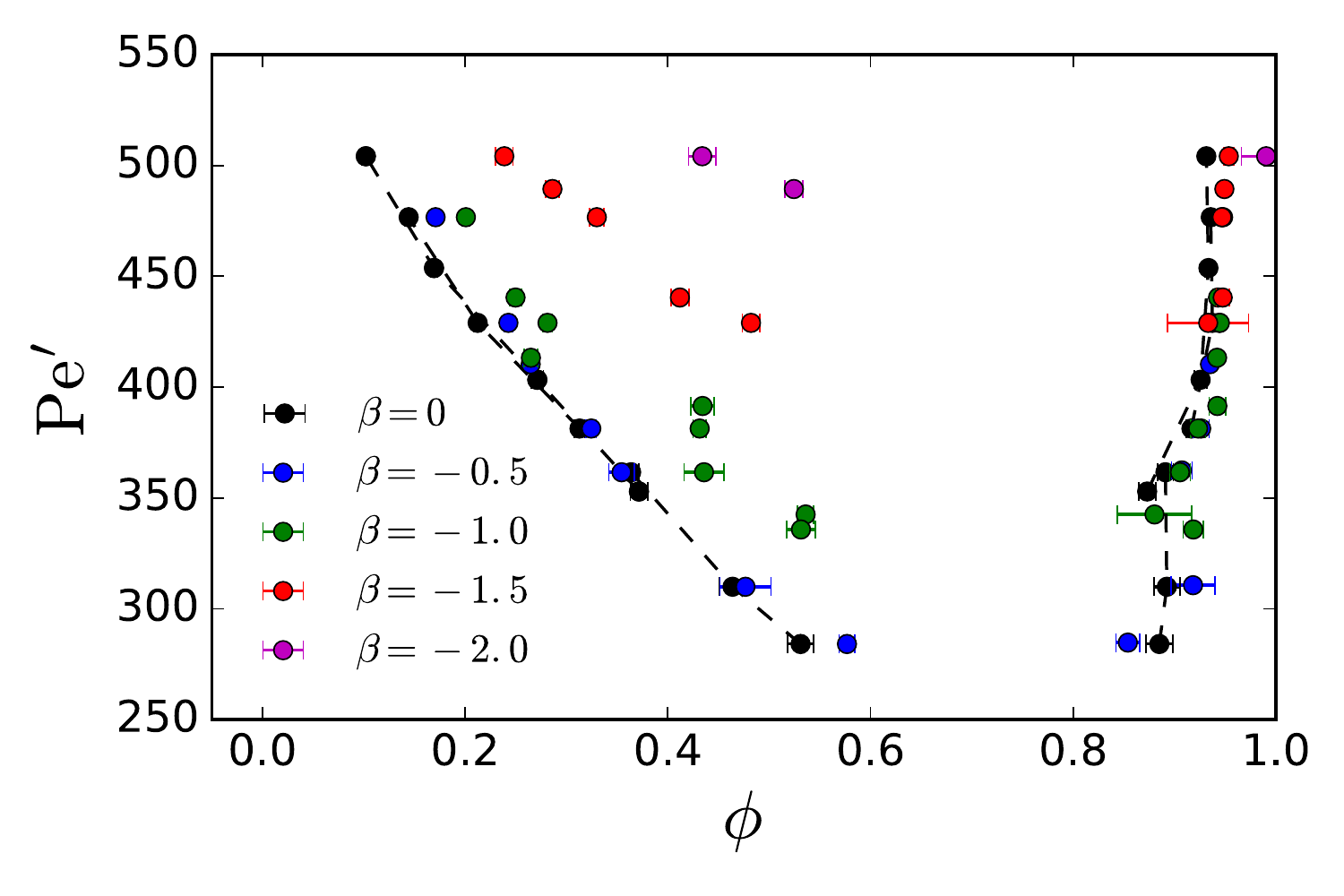}
	\includegraphics[width=0.5\textwidth]{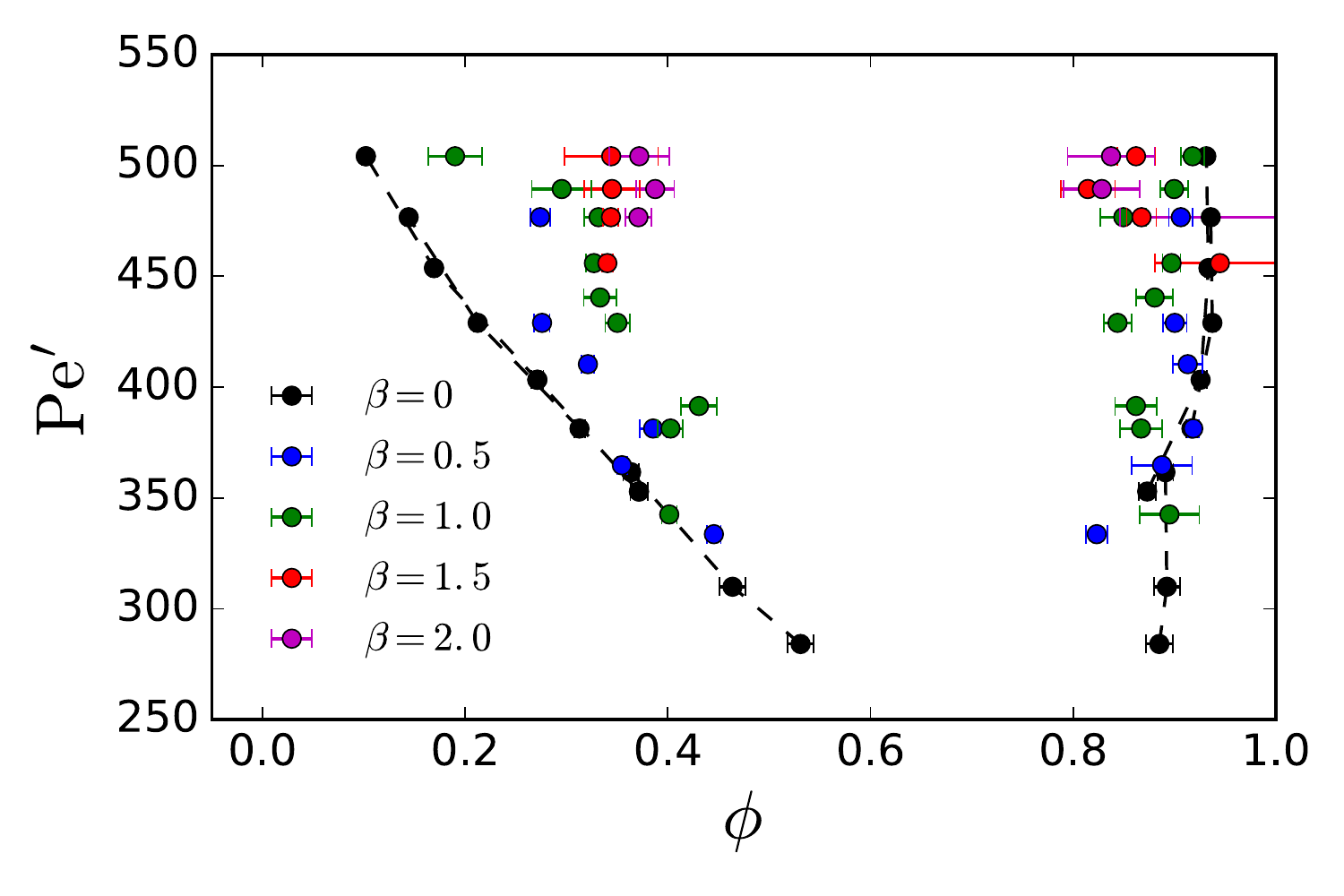}
	\caption{Binodal lines of pushers (top panel) and pullers (bottom panel) for different dipole strengths $\beta$. 
	For reference, the data from Fig.\ \ref{fig:coex}:\ \emph{bottom} is shown as black.
	\label{fig:beta}}
\end{figure}

The flow field around squirmers is strongly controlled by the dipole strength $\beta$ used in our swimmer model. 
For $\beta \neq 0$, the flow field of a single squirmer in a bulk fluid is reminiscent of pushers ($\beta<0$) and pullers ($\beta>0$) \cite{Spagnolie:2012gk,Zottl:2016kr}.
It has a strong effect on how swimmers reorient due to nearby surfaces or other squirmers as demonstrated in Refs. \cite{Spagnolie:2012gk,Zottl:2016kr,Zottl:2014fn}. 
The qualitative effect of increasing $|\beta|$ has been shown in our earlier work \cite{Zottl:2014fn} to largely suppress phase separation, requiring systems of increasing $\bar{\phi}$ in order for stable clusters to form.

The binodal lines for different $\beta$ are compared to the neutral-swimmer case in Fig.~\ref{fig:beta} for pushers (top panel) and pullers (bottom panel).
The overall trend is that for increasing $| \beta |$ the coexistence regime becomes smaller. 
For pushers, one realizes very nicely how the binodals are mainly shifted up, thereby increasing the critical P\'eclet number.
The same applies to pullers, however, the gas binodals are also deformed compared to $\beta = 0$. 
We observe that puller squirmers are more mobile in the dense phase compared to pushers. 
Therefore, the corresponding binodals are shifted to smaller $\phi$.

For pushers and pullers in full three dimensions spontaneous polar order was observed for small but non-zero $\beta$\cite{Alarcon:2013dg}.
This results in a net motion of swimmers along a common direction.
We do not find such a net motion for small $\beta$ in our simulations.
This might be due to the fact that in our quasi-two-dimensional geometry the far fields of neutral and non-neutral squirmers both decay as $1/r^{2}$ \cite{Brotto:2013fj,Delfau:2016cv}.
Thus, squirmers with small $\beta\neq 0$ do not have a sufficiently different far field compared to neutral squirmers.

\subsection{Melting of the Cluster Phase}

\begin{figure}
	\includegraphics[width=0.5\textwidth]{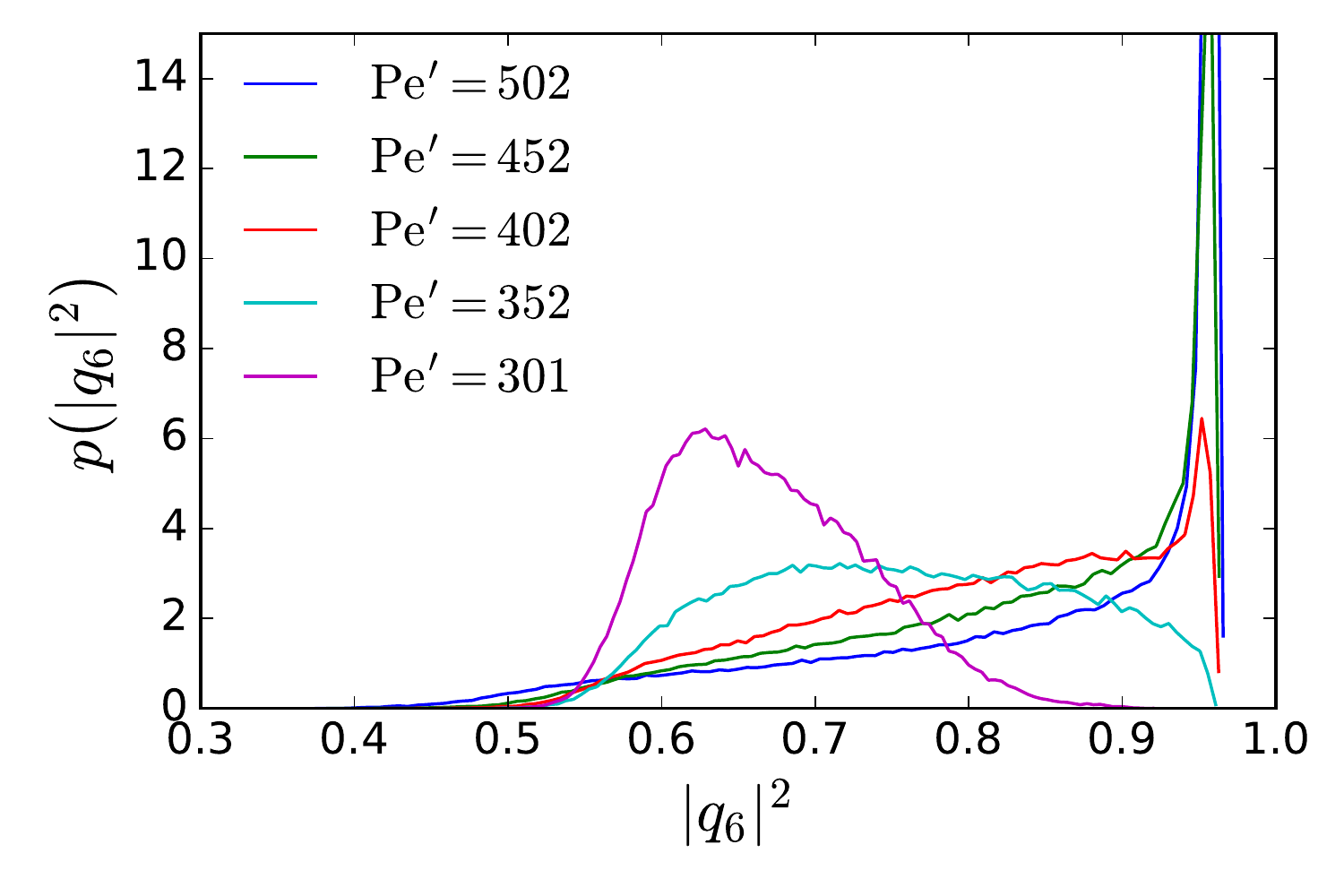}
	\caption{Probability distribution of the six-fold bond parameter $|q_6|^2$ within the cluster phase for $\bar{\phi}=0.64$ and a range of effective P\'eclet numbers $\mathrm{Pe}^{\prime}$. \label{fig:melting}}
\end{figure}

In the region below the critical P\'eclet number in Fig.\ \ref{fig:coex}:\ \emph{bottom} the phase is isotropic. 
On the other hand, at large effective P\'eclet numbers we observe hexagonal order in the dense phase, which hence has to undergo a melting transition if the Peclet number is reduced. 
It is not our intention to present a thorough discussion of this subtle phenomenon but to only give an indication for melting.

The dense branch of the binodal lines of neutral squirmers and of pushers (see Fig.\ \ref{fig:beta}: \emph{top}) reveal a similar trend.
At $\mathrm{Pe}^\prime \approx 400$ and below, the density of the dense phase decreases and stabilizes at a lower value around $\mathrm{Pe}^\prime \approx 350$.
In particular, this feature is clearly seen for neutral squirmers and coincides with a loss of hexagonal order in the cluster.
As in Ref.\ \cite{Zottl:2014fn} we introduce the bond parameter $|q_6|^2$ to measure local sixfold bond-orientational order around a squirmer.
In Fig~\ref{fig:melting} we plot the probability distribution for $|q_6|^2$ in the dense phase for a range of $\mathrm{Pe}^\prime$.
For $\mathrm{Pe}^\prime \geq 402$ a peak at $|q_6|^2 \approx 1$ dominates the distribution and indicates hexagonal order in the dense phase.
The hexagonal domains are stable over time with only short perturbations when defects and vacancies move through the hexagonal lattice due to the random component of the squirmers' motions.
These perturbations slightly reduce the time-averaged $|q_6|^2$ and shift the peak position below 1.

Below $\mathrm{Pe}^\prime \geq 402$ the sharp peak in the distribution function disappears indicating a melting transition with a loss of hexagonal order.
In the simulations we still observe stable clusters, which now form fluid droplets. During the melting transition we also
observe an increase of the mobility of the squirmers by a factor of 5-10, when monitoring the mean-square displacement in the
dense phase.
The supplemental material$^\dag$ contains videos M5 and M6 showing the hexagonal and fluid dense phase, respectively. 
In video M6 we see that, although hexagonal patches exist, they are not stable over time.
Finally, when lowering $\mathrm{Pe}^\prime$ towards the critical point, the fluid cluster ``evaporates'' as the system exits the 
phase-coexistence regime.

\section{Summary and Conclusion}

We have studied the phase separation of model hydrodynamic microswimmers called squirmers in quasi-two-dimensional confinement extending our previous investigations \cite{Zottl:2014fn}.
The full three-dimensional hydrodynamics was simulated in order to accurately capture the squirmer interactions.
This makes our simulations a realistic representation of real-world artificial and biological microswimmers.
By employing a parallelized implementation of the MPCD algorithm on over 1008 CPU cores, we were able to simulate thousands of squirmers, allowing us to quantitatively determine the binodal lines for different mean densities $\bar{\phi}$ and squirmer parameters $\beta$.
The coarsening dynamics towards the phase-separated system shows a characteristic intermediate diffusive regime and a final rather ballistic regime due to compactification of the dense squirmer cluster. 
However, most strikingly, we found that the position of the binodal lines is strongly influenced by $\bar{\phi}$. 
This effect has not been observed for active Brownian particles, which only interact sterically, and thus is of real hydrodynamic origin.
We were able to collapse the different binodals on a single master curve by introducing an effective P\'eclet number.
Extending the mechanical pressure balance of Ref.\ \cite{Bialke:2015cq,Takatori:2014do} by a hydrodynamic pressure due to the squirmer flow field, the reason for the different binodals became apparent.

We also explored the dependence of the binodal line on $\beta$, a parameter which allows us to tune the swimmer type and their flow fields between pushers and pullers.
We found that in all cases going to strong pushers or pullers suppresses phase separation by shifting the phase-coexistence region to larger P\'eclet numbers.
The shape of the coexistence region does, however, depend on whether the squirmers are pushers or pullers.

Finally, we examined the structure of the dense phase as the squirmers' swimming speed, the P\'eclet number, is reduced. 
We find that fast squirmers in the dense phase form a hexagonal cluster, which melts with decreasing P\'eclet number into a fluid cluster. 
This is expected since close to the critical P\'eclet number the swimmer fluid is isotropic.

\section{Acknowledgements}

We would like to thank I. Aranson, R. Golestanian, G. Gompper, R. Kapral, K. Kroy, T. Speck, and J. Tailleur for stimulating discussions and the referees for their very helpful feedback.
This project was funded under the DFG priority 
program SPP 1726 ``Microswimmers - from Single Particle 
Motion to Collective Behaviour'' (grant number STA352/11). Simulations were conducted at the ``Norddeutscher Verbund f\"ur 
Hoch- und H\"ochstleistungsrechnen'' (HLRN), project number bep00050.

\section{Appendix}
\label{sec:appendix}

\subsection{Position-Resolved Time Averaging}

\begin{figure}
	\centering
	\includegraphics[width=0.4\textwidth]{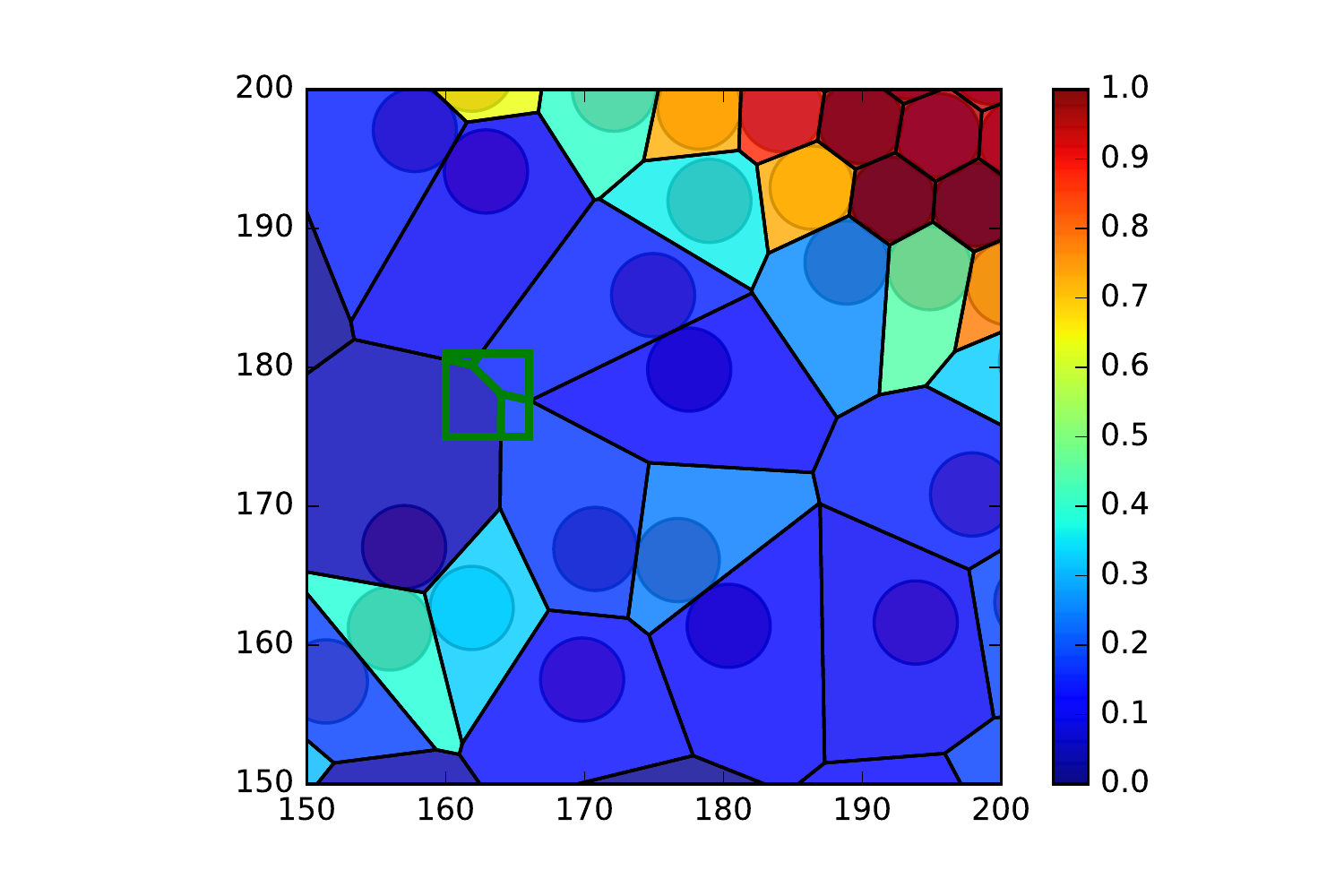}
	\caption{Illustration for the spatially-resolved time average of local density.
	Each Voronoi cell with area $A_i$ is assigned a local density $\phi_i = 1/A_i$ (colorbar).
	The average density at a pixel (green square) is the weighted average of the local density in all overlapping Voronoi cells, based on the degree of overlap. \label{fig:voronoi}}
\end{figure}

In order to associate an average quantity to a position, one commonly divides up the system into a grid. 
We call each cell of the grid a ``pixel''.
A spatially-resolved average can thus be found by simply averaging over the squirmers in each pixel.
Since we have only a few thousand squirmers, each pixel would need to be relatively large in order to capture enough squirmers per pixel.
This would prevent the accurate resolution of the phase interface.

We will use the local density to illustrate our method. 
First we partition the system into its Voronoi cells.
Each Voronoi cell contains a single squirmer, and thus its local density is $\phi_i = 1/A_i$ where $A_i$ is the area of the $i$-th Voronoi cell.
Our grid of pixels is then assigned a local density
\begin{align}
	\phi(\vv{r}_{i,j}) = \sum_{k\;\in\;\mathrm{overlap}} \phi_k w_k
\end{align}
where $\vv{r}_{i,j}$ is the position of the pixel with indices $i$ and $j$. The sum is taken over those Voronoi cells overlapping with the square boundaries of the pixel. $\phi_k$ is the local density of Voronoi cell $k$, and $w_k=a_\mathrm{overlap}/a_\mathrm{pixel}$ is a weight equal to the Voronoi cell's overlap area $a_\mathrm{overlap}$ with the pixel normalized by the pixel area $a_\mathrm{pixel}$. An example of this is shown in Fig.~\ref{fig:voronoi}.

All time averages are taken with respect to a specific spatial position.
Due to the motion of the squirmers, the time averages of $\phi(\vv{r}_{i,j})$ eventually wash out the structure of the voronoi construction, leaving only those density inhomogeneities that are stationary in space.
We use the same construction for calculating local pressure and $|q_6|^2$ values.

\subsection{Comparison with Active Brownian Particles}
\label{sec.comp}

\begin{figure}
	\centering
	\includegraphics[width=0.45\textwidth]{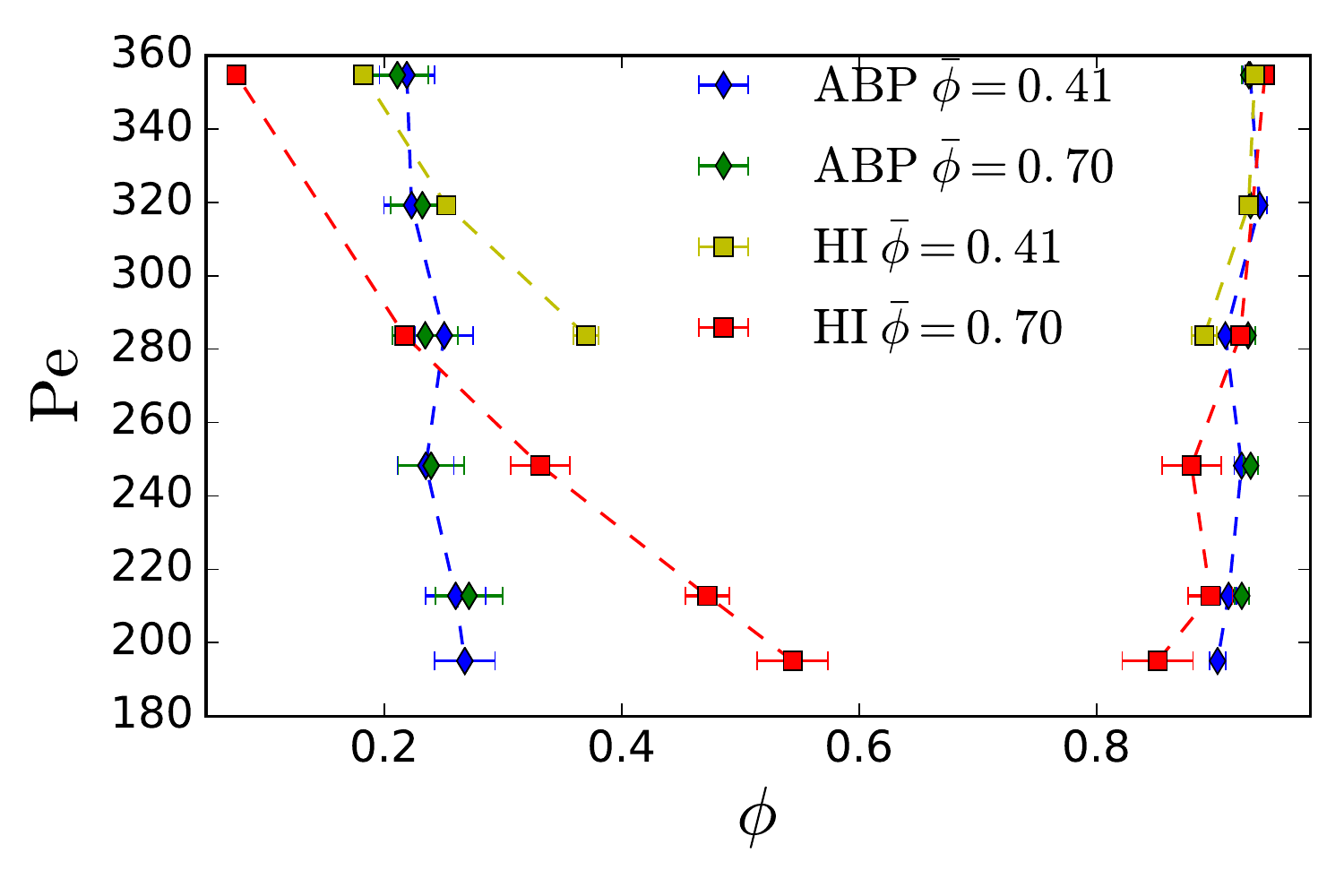}
	\caption{Comparison between binodals for active Brownian particles (ABP) and the binodals for squirmers (HI) from Fig.~\ref{fig:coex}. Mean densities and P\'eclet numbers, as well as simulation box size and squirmer radius correspond to the range used for the full hydrodynamic simulations.
	\label{fig:binodal-abp}}
\end{figure}

Figure~\ref{fig:binodal-abp} compares the binodals from Fig.~\ref{fig:coex} with those for active Brownian particles in quasi two dimensions. 
Brownian dynamics simulations where conducted for the same range of parameters and the same geometry as the full hydrodynamic simulations.
Similarly to simulations conducted in two dimensions\cite{Speck:2015gc} for the range of P\'eclet numbers simulated, the coexistence densities for active Brownian particles did not vary much. 
We also note that the binodals for two different mean densities $\bar{\phi} = 0.41$ and $\bar{\phi}=0.70$ lie on top of each other.
For comparison, the binodals for the hydrodynamically interacting (HI) squirmers from Fig.~\ref{fig:coex} are shown.

\subsection{Position-Resolved Swim and Steric Pressure}
\label{app.pressure}

\begin{figure}
	\centering
	\includegraphics[width=0.4\textwidth]{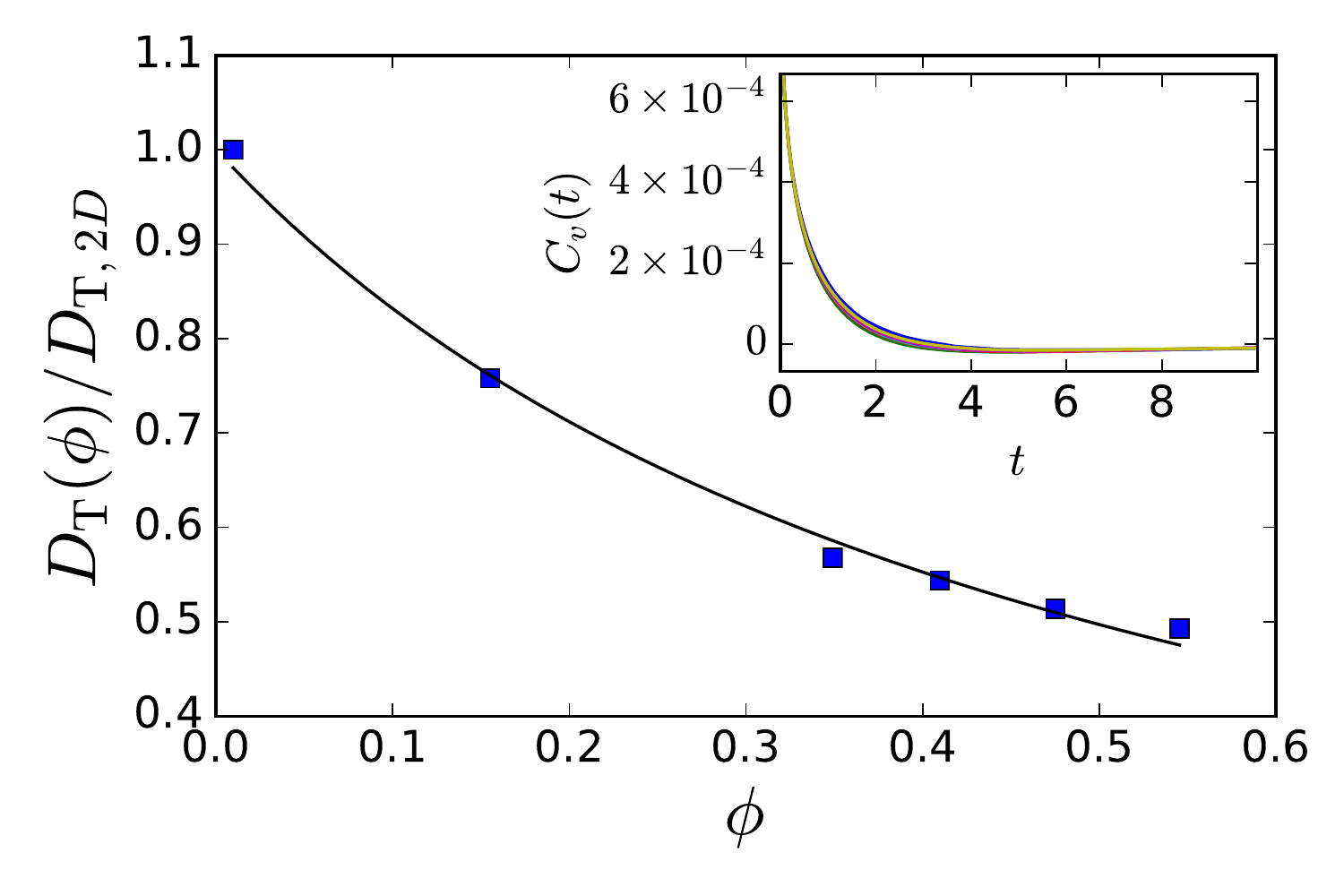}
	\caption{Self-diffusion coefficient of passive particles plotted versus their density $\phi$ confined between two walls with distance $H=8R/3$. 
	The full line is a fit to $D_{T,2D} (\phi) = D_{T,2D} (\phi =0) / (1+b\phi)$ with $b=2.027$.
	Inset: The corresponding in-plane velocity-auto-correlation functions $C_v(t) = \langle \vv{v}(t)\cdot\vv{v}(0) \rangle/2k_\mathrm{B}T$ for the different densities in the main plot. The differences are rather subtle. Time is given in MPCD units.
	\label{fig.selfdiff}}
\end{figure}

\begin{figure}
	\centering
	\includegraphics[width=0.3\textwidth]{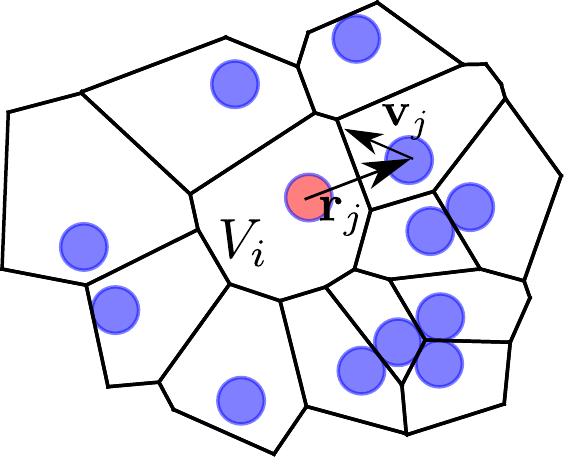}
	\caption{Illustration of the swim pressure in the central Voronoi cell (central particle colored in red).
	The coordinate system is centred at the central particle. 
	Its Voronoi cell has a volume $V_i$.
	Nearby particles exert a swim pressure
	$\gamma_0 v_0 \langle \sum_j \hat{\vv{e}}_j \cdot \vv{r}_j \rangle/3V_i$	``on'' the Voronoi cell of the central particle, where $j$ runs over all neighboring Voronoi cells.
	\label{fig:Voroni-p_act}}
\end{figure}

Before we explain, how we calculate the pressure values, we present in the inset of Fig.\ \ref{fig.selfdiff} the velocity-auto-correlation function for the in-plane velocity components of a passive particle, determined in MPCD simulations for different mean densities of passive particles. 
The  self-diffusion coefficient $D_{T,2D} (\phi) = \int \langle \mathbf{v}(t) \cdot \mathbf{v}(0) \rangle \dd t/2 k_\mathrm{B}T$ is then plotted in the main graph of  Fig.\ \ref{fig.selfdiff} 
and fitted by $D_{T,2D} (\phi) = D_{T,2D} (\phi =0) / (1+2.027\phi)$. Using the Einstein relation, we obtain a density-dependent
friction coefficient, $\gamma_0(\phi) = k_\mathrm{B}T /D_{\mathrm{T}, 2D}(\phi)$.

Now, we apply Eq.\ (\ref{eq:p_act}) to the individual Voronoi cells and its surround neighbours (\emph{cf.}\ Fig.\ \ref{fig:Voroni-p_act}).
Our system volume now is $V_i \approx HA_i$, \emph{i.e.}, the height of the system $H$ times the 2D-Voronoi cell area $A_i$, and the neighbours are the immediate neighbours in terms of the Voronoi cell.
On this level, the force generating the active motion $\vv{F}_\mathrm{swim}= \gamma_0 \vv{v}_j$ of neighbouring squirmers will act on the central
Voronoi cell, \emph{i.e.} inducing a pressure on its ``walls''. 
There is an additional subtlety here: the position vectors $\mathbf{r}_j$ of the neighboring squirmers always have to start from the central swimmer, for which local pressure is calculated,
otherwise shear terms would appear in the derivation of Eq.~\eqref{eq:p_act} \cite{Speck:2016jl}.
Furthermore, the mean velocity also needs to be subtracted from the individual swimmer velocities.
Thus we are performing calculations in the centre-of-mass frame of the neighbourhood of each squirmer.
We reduce the amount of fluctuations by taking the time average for each pixel (\emph{cf.} previous subsection on spatial averaging).
This time average is longer than the decorrelation time $D_{\mathrm{R}}^{-1}$ of the swimmers as required by Eq.~\eqref{eq:p_act}.

Similarly, we use the Voronoi cells to also introduce a local steric pressure following Ref.\ \cite{Tsai:1979dj}.
The steric forces acting on one squirmer include forces from its neighbors and also from the bounding walls.

\balance


\bibliography{rsc_art_arxiv} 
\bibliographystyle{rsc} 

\end{document}